\documentclass[preprint,12pt,pre]{revtex4}%
\usepackage{amsfonts}
\usepackage{amsmath}
\usepackage{amssymb}
\usepackage{graphicx}
\usepackage{theorem}%
\providecommand{\U}[1]{\protect\rule{.1in}{.1in}}
\newtheorem{theorem}{Theorem}

\begin{document}
\preprint{UATP/1005}
\title{Non-equilibrium thermodynamics.III. Thermodynamic Principles, Entropy
Continuity during Component Confinement, Energy Gap and the Residual Entropy }
\author{P. D. Gujrati}
\email{pdg@uakron.edu}
\affiliation{Departments of Physics, Department of Polymer Science, The University of
Akron, Akron, OH 44325}

\begin{abstract}
To investigate the consequences of component confinement such as at a glass
transition and the well-known energy or enthalpy gap (between the glass and
the perfect crystal at absolute zero, see text), we follow our previous
approach [Phys. Rev. E \textbf{81}, 051130 (2010)] of using the second law
applied to an isolated system $\Sigma_{0}$ consisting of the homogeneous
system $\Sigma$ and the medium $\widetilde{\Sigma}$. We establish on general
grounds the continuity of the Gibbs free energy $G(t)$ of $\Sigma$ as a
function of time at fixed temperature and pressure of the medium. It
immediately follows from this and the observed continuity of the enthalpy
during component confinement that the entropy $S$ of the open system $\Sigma$
must remain continuous during a component confinement such as at a glass
transition. We use these continuity properties and the recently developed
non-equilibrium thermodynamics to formulate thermodynamic principles of
additivity, reproducibility, continuity and stability that must also apply to
non-equilibrium systems in internal equilibrium. We find that the
irreversibility during a glass transition only justifies the residual entropy
$S_{\text{R}}$ to be at least as much as that determined by disregarding the
irreversibility, a common practice in the field. This justifies a non-zero
residual entropy $S_{\text{R}}$ in glasses, which is also in accordance with
the energy or enthalpy gap at absolute zero. We develop a statistical
formulation of the entropy of a non-equilibrium system, which results in the
continuity of entropy during component confinement in accordance with the
second law and sheds light on the mystery behind the residual entropy, which
is consistent with the recent conclusion [Symmetry \textbf{2}, 1201 (2010)]
drawn by us.

\end{abstract}
\date[September 28, 2010]{}
\maketitle

\section{Introduction\label{Marker_Introduction}}

\subsection{Irreversibility\label{Marker_Irreversibility}}

\begin{figure}
[ptb]
\begin{center}
\includegraphics[
height=2.5322in,
width=5.2243in
]%
{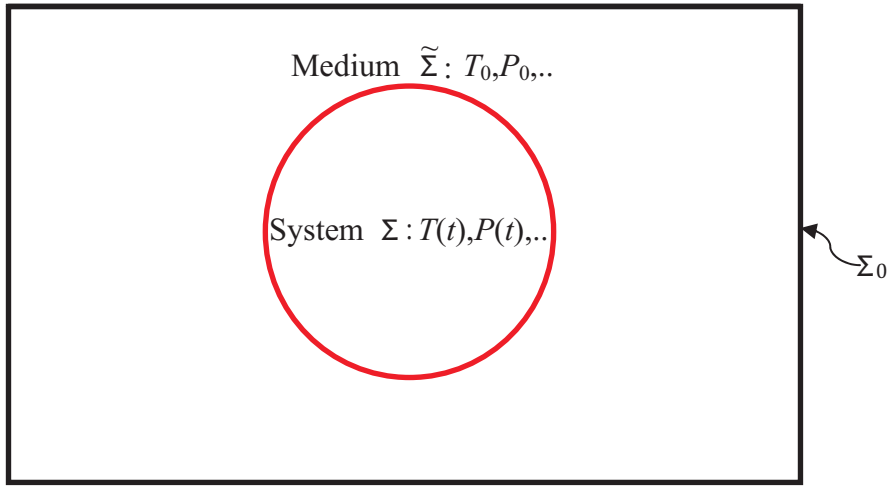}%
\caption{Schematic representation of a system $\Sigma$ and the medium
$\widetilde{\Sigma}$ surrounding it to form an isolated system $\Sigma_{0}$.
The medium is described by its fields $T_{0},P_{0},$ etc. while the system, if
in internal equilibrium (see text) is characterized by $T(t),P(t),$ etc.}%
\label{Fig_Systems}%
\end{center}
\end{figure}

In the previous two papers \cite{Guj-NE-I,Guj-NE-II}, to be referred to as I
and II respectively here, we initiated a general statistical mechanical
investigation of non-equilibrium systems undergoing irreversible changes
during approach to their equilibrium state. It is a very general approach that
is not restricted to glasses alone but covers all non-equilibrium situations.
In I, we deal with homogeneous systems, while in II we deal with inhomogeneous
systems. The approach is truly statistical mechanical in nature, and is based
on applying the second law of thermodynamics to an extremely large isolated
system, which we denote by $\Sigma_{0}$; it\ consists of the macroscopic
system $\Sigma$ of interest in a medium denoted by $\widetilde{\Sigma}$
containing it; see Fig. \ref{Fig_Systems}. We will use \emph{body} in this
work to refer to any of the three systems. According to the second law, the
entropy $S_{0}$ of an isolated system $\Sigma_{0},$ which is a sum of the
entropies $S$ and $\widetilde{S}$ of the system and the medium, respectively,%
\begin{equation}
S_{0}(t)=S(t)+\widetilde{S}(t), \label{Entropies_Sum}%
\end{equation}
can never decrease in time
\cite{Gibbs,Becker,Huang,Rice,Landau,Gujrati-Symmetry}:%
\begin{equation}
\frac{dS_{0}(t)}{dt}\geq0. \label{Second_Law}%
\end{equation}
Any change in the entropy of the isolated system is due to irrevesibility, so
that $d_{\text{i}}S_{0}\equiv dS_{0}\geq0$, and $d_{\text{e}}S_{0}\equiv0$ in
the standard notation \cite{Donder,deGroot,Prigogine,Guj-NE-I,Guj-NE-II}. Even
for a body, which is not isolated, $d_{\text{i}}S\equiv dS-d_{\text{e}}S\geq
0$, with $d_{\text{e}}S_{0}\neq0$; indeed, the latter can be of any sign.
%TCIMACRO{\FRAME{ftbpFU}{5.2243in}{2.5322in}{0pt}{\Qcb{Schematic representation
%of a system $\Sigma$ and the medium $\widetilde{\Sigma}$ surrounding it to
%form an isolated system $\Sigma_{0}$. The medium is described by its fields
%$T_{0},P_{0},$ etc. while the system, if in internal equilibrium (see text) is
%characterized by $T(t),P(t),$ etc.}}{\Qlb{Fig_Systems}}{system_modified_1.eps}%
%{\special{ language "Scientific Word";  type "GRAPHIC";
%maintain-aspect-ratio TRUE;  display "USEDEF";  valid_file "F";
%width 5.2243in;  height 2.5322in;  depth 0pt;  original-width 5.1673in;
%original-height 2.4915in;  cropleft "0";  croptop "1";  cropright "1";
%cropbottom "0";  filename 'System_Modified_1.eps';file-properties "XNPEU";}}}%
%BeginExpansion
%EndExpansion
Two important observations about the entropy $S_{0}$ are in order for Eq.
(\ref{Second_Law}) to have a content.:

\begin{enumerate}
\item The entropy $S_{0}$ \emph{exists} under all conditions; thus it includes
the situation when $\Sigma_{0}$ is not in equilibrium or when $\Sigma$
undergoes a glass transition.

\item The entropy $S_{0}$ is a \emph{continuous function} of all its arguments
$\mathbf{Z}_{0}(t)$ (see below), not shown in Eq. (\ref{Second_Law}), and the
time $t$, which is shown.
\end{enumerate}

What happens inside the isolated system (loss of ergodicity in parts of
$\Sigma_{0}$ due to component confinement, chemical reactions, phase changes,
turbulence, viscous deformation inside it, etc.) cannot affect the direction
of the inequality, which makes it the most general principle of
non-equilibrium thermodynamics
\cite{Donder,deGroot,Prigogine,Guj-NE-I,Guj-NE-II}. All the above entropies,
in addition to being an explicit function of time $t$, are also functions of
extensive observables and internal variables (see II for more details) that
themselves may vary with time. Of the observables, the energy, volume and the
number of particles are the most important extensive observables from an
experimental point of view. In the sequel to the above two paper I and II, we
focus on non-equilibrium thermodynamic principles dictating the behavior of a
\emph{homogeneous} system $\Sigma$, the general way its non-equilibrium
entropy should behave during relaxation, the way $S$ is determined in terms of
its microstates, and the way $S$ must behave when the system becomes confined
to a component of the phase space \cite{Palmer,Jackle,Gujrati-book} during
some interval of time (for example, at a glass transition), and the concept of
the \emph{residual entropy}, the entropy that $\Sigma$ will have at absolute
zero. As our interest is in investigating the out of equilibrium properties of
$\Sigma$, we will assume throughout this work and as was done in I and II that
the medium is in internal equilibrium at all times. As a consequence,
$\widetilde{\Sigma}$ has a well-defined \emph{constant} temperature $T_{0},$
pressure $P_{0}$ and other fields and affinities (see below) at all times. In
contrast, the system may have no well-defined temperature, pressure, etc.
unless it is also in internal equilibrium, in which case, we will denote their
instantaneous values by $T(t),P(t),$ etc.; see I for details. They are usually
different from the those of the medium unless equilibrium is reached, which
occurs as $t\rightarrow\infty$. It was shown in I that $T(t)\geq T_{0}$ in a
cooling and $T(t)\leq T_{0}$ in a heating experiment, and that
\begin{equation}
dQ=T(t)dS(t)\equiv T_{0}d_{\text{e}}S(t), \label{Heat_Entropy Relation}%
\end{equation}
where $d_{\text{e}}S(t)$ represents the entropy exchange with the medium; the
irreversible entropy generation or the uncompensated transformation of
Clausius and de Donder \cite{Donder,deGroot,Prigogine} within the system is
represented by $d_{\text{i}}S(t)\equiv dS(t)-d_{\text{e}}S(t)\geq0$.

\subsection{Determination of Entropy}

The entropy can now be calculated in principle by using Eq.
(\ref{Heat_Entropy Relation}) even when the process is irreversible. As the
experimentalist has a control over $T_{0},P_{0},$ the entropy can be
calculated by varying $T_{0}$, keeping $P_{0}$ fixed. The heat capacity
$C_{P}$ at constant $P_{0}$ in terms of the enthalpy $H(t)$ and the volume
$V(t)$ is given by
\begin{equation}
C_{P}(t)\equiv\left(  \frac{\partial H(t)}{\partial T_{0}}\right)  _{P_{0}%
}+[P(t)-P_{0}]\left(  \frac{\partial V(t)}{\partial T_{0}}\right)  _{P_{0}},
\label{Heat Capacity}%
\end{equation}
see I, so that
\begin{equation}
dS(t)=\frac{C_{P}}{T(t)}dT_{0}\left\{
\begin{array}
[c]{c}%
\leq C_{P}dT_{0}/T_{0}\ \ \ \text{in cooling}\\
\geq C_{P}dT_{0}/T_{0}\ \ \ \text{in heating}%
\end{array}
\right.  , \label{Entropy_HeatCapacity_Bounds}%
\end{equation}
and can be used to obtain the entropy using the standard integration process%
\begin{equation}
S(T_{0})=S(0)+%
%TCIMACRO{\tint \limits_{0}^{T_{0}}}%
%BeginExpansion
{\textstyle\int\limits_{0}^{T_{0}}}
%EndExpansion
\frac{C_{P}dT_{0}}{T(t)}+\Delta S_{\text{C}}\left\{
\begin{array}
[c]{c}%
\leq S(0)+%
%TCIMACRO{\tint \limits_{0}^{T_{0}}}%
%BeginExpansion
{\textstyle\int\limits_{0}^{T_{0}}}
%EndExpansion
C_{P}dT_{0}/T_{0}+\Delta S_{\text{C}}\ \ \ \text{in cooling}\\
\geq S(0)+%
%TCIMACRO{\tint \limits_{0}^{T_{0}}}%
%BeginExpansion
{\textstyle\int\limits_{0}^{T_{0}}}
%EndExpansion
C_{P}dT_{0}/T_{0}+\Delta S_{\text{C}}\ \ \ \text{in heating}%
\end{array}
\right.  , \label{Entropy_HeatCapacity}%
\end{equation}
where $S(0)$ is the entropy at absolute zero, and $\Delta S_{\text{C}}$
represents the sum of all possible discontinuities over the range $\left(
0,T_{0}\right)  $ mainly due to latent heats at possible first-order
transitions; it also includes any possible entropy discontinuity due to
component confinement. The above inequalities are a simple extension of is a
well-known classic result \cite{note0}, which has been rederived in our
approach. Because of a non-negative heat capacity, the entropy is a
\emph{monotonic} \emph{increasing} function of the temperature:%
\begin{equation}
\frac{\partial S}{\partial T_{0}}>0 \label{Monotonic_S}%
\end{equation}
for all finite temperatures. The presence of an internal order parameter can
also be accounted for as we will discuss later. We find that the integral
\begin{equation}%
%TCIMACRO{\tint \limits_{0}^{T_{0}}}%
%BeginExpansion
{\textstyle\int\limits_{0}^{T_{0}}}
%EndExpansion
C_{P}dT_{0}/T_{0}>0 \label{Heat_Capacity Integral}%
\end{equation}
provides an upper bound to the entropy $S(t)$ of the system during a cooling
experiment, and a lower bound for it during a heating experiment, when the
system is out of equilibrium. The difference in the two bounds, which we will
denote by $\Delta S_{\text{B}}(T_{0},P_{0})$ here, strongly depends on the
rates of cooling and heating. The two bounds are exactly identical so that
$\Delta S_{\text{B}}(T_{0},P_{0})\equiv0$, when the system is in equilibrium.
Therefore, $\Delta S_{\text{B}}(T_{0},P_{0})$ can be used as a measure of the
\emph{irreversibility} when no internal variables (order parameters) are
invoked, as was recently done by Johari and Khouri \cite{Johari-New}. However,
introducing internal variables allows us to obtain the entropy without any
error, as we will discuss here, even when the system is out of equilibrium and
irreversibility is present so that the thermodynamic entropy can be always
calculated, at least in principle. Irreversible entropy generation appears
through the internal order parameter in the calculation. This is a
well-established approach to non-equilibrium thermodynamics
\cite{Donder,deGroot,Prigogine}; see also \cite{Guj-NE-I,Guj-NE-II}, which we
carefully discuss in Sect. \ref{Marker_Useful_Results} for the sake of continuity.

\subsection{Non-equilibrium Thermodynamic Principles}

We have a through understanding of thermodynamic principles (additivity,
reproducibility, uniqueness and stability, all defined later) and the
statistical entropy for an equilibrium system
\cite{Gibbs,Becker,Huang,Rice,Landau}. The situation is not so obvious when
the system is out of equilibrium. We wish to examine the conditions necessary
for the applicability of these principles to, and obtain the statistical
formulation of the entropy of, a non-equilibrium system with the aim to
develop a non-equilibrium statistical thermodynamics. While in traditional
non-equilibrium thermodynamics
\cite{Donder,deGroot,Prigogine,Guj-NE-I,Guj-NE-II}, the entropy of any body is
postulated to exist under the assumption of local equilibrium, its actual
value and its statistical connection with microstate probabilities is of no
interest as only the difference in the entropy has any meaning. The same is
also true of the thermodynamics applied to glasses
\cite{GoldsteinSimha,Jones,Nemilov-Book,Gutzow-Book,Gujrati-book}. The actual
value of $S$ is normally set by imposing additional requirement such as by
invoking Nernst postulate of the third law
\cite{Landau,Gujrati-Nernst,Gujrati-Fluctuations}. Our interest here is
different. We wish to identify the entropy of a body when it is out of
equilibrium using statistical concepts. It should be stated here that the
entropy for a body in equilibrium is well understood in equilibrium
statistical mechanics \cite{Gibbs,Huang,Rice,Landau}. Thus, any new attempt to
define the entropy for a non-equilibrium state must reduce the the well-known
form of equilibrium entropy.

\subsection{Component confinement, Entropy Reduction and the Residual Entropy}

\subsubsection{Component Confinement and the Gap}

\begin{figure}
[ptb]
\begin{center}
\includegraphics[
height=3.8649in,
width=5.3264in
]%
{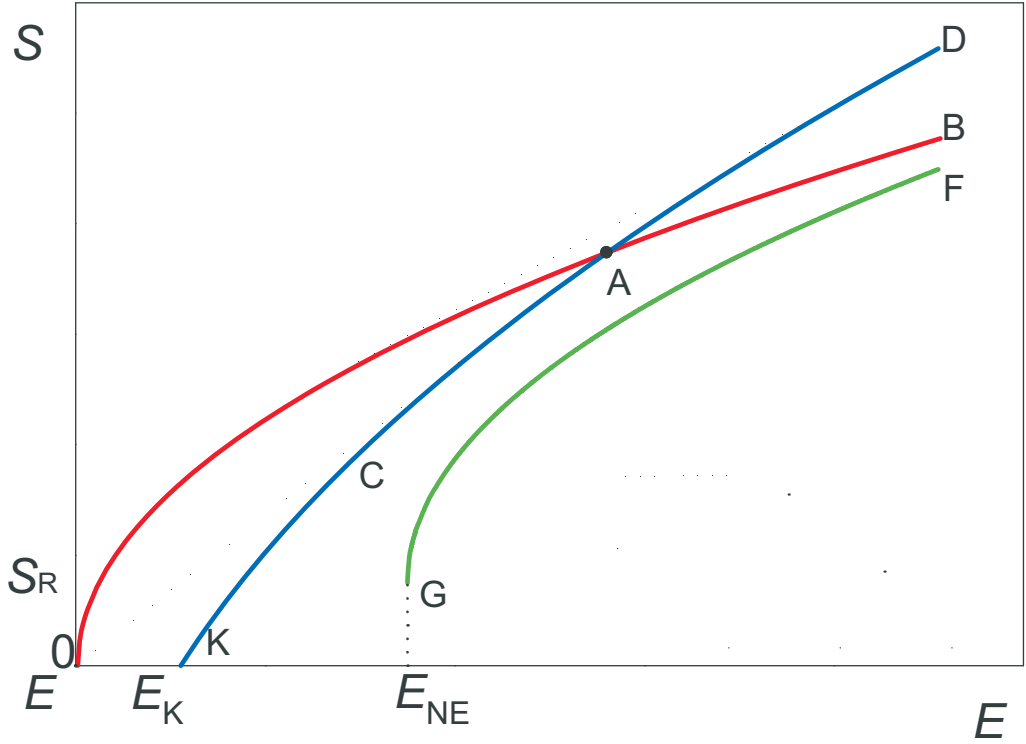}%
\caption{Schematic form of the communal entropy DACK of the disordered
liquid/supercooled liquid and the entropy BA0 of the corresponding ordered
crystalline state as a function of the energy; other extensice observables are
held foxed. The common tangent to DACK and BA0 gives the inverse melting
temperature $T_{\text{M}}$. The entropy of some non-equilibrium disordered
state for some non-zero cooling rate $r$ is given by FG, which must strictly
lie below in accordance with the second law. It terminates at G at an energy
$E_{\text{NE}}$, which is strictly higher than the energy (taken to be $0$
here) of the pure crystal at absolute zero or the energy $E_{\text{K}}$ at K
on DACK. The entropy at G gives the residual entropy $S_{\text{R}}$. Even
though we have shown FG disconnected from DACK for some arbitrary
non-equilibrium state, it continuously emerges out of the latter at some point
C at a temperature below $T_{\text{M}}$ in a glass transition$.$The portion CG
in this case represents the entropy of the glassy state. As the cooling rate
becomes slower, CG moves towards CK until it finally merges with it as
$r\rightarrow0$. The terminal point K is known as the \emph{ideal glass
transition} where the entropy of the equilibrium supercooled liquid vanishes.
As FG moves towards ACK, $E_{\text{NE}}$ moves towards the ideal glass energy
$E_{\text{K}}$ and the residual entropy continues to decrease until the latter
finally vanishes in the ideal glass transition at K. However, for any $r>0$,
$S_{\text{R}}$ remains non-zero due to the presence of excitations in the
ideal glass that raises its energy above $E_{\text{K}}$. }%
\label{Fig_Gap_Model}%
\end{center}
\end{figure}

Phase space confinement at a phase transition such as a liquid-gas transition
is a well-known phenomenon in equilibrium statistical mechanics
\cite{Huang,Landau,GujratiResidualentropy,Gujrati-Symmetry}. The component
confinement also occurs when the system undergoes symmetry breaking such as
during magnetic transitions, crystallizations, etc. In such cases, the
confinement occurs under equuilibrium conditions, and is well understood. The
confinement in phase space can also occur under non-equilibrium conditions,
when the observational time scale $\tau_{\text{obs}}$\ becomes shorter than
the equilibration time of the system, such as for glasses
\cite{GoldsteinSimha,Jones,Nemilov-Book,Gutzow-Book,Gujrati-book}, whose
behavior and properties have been extensively studied; see Fig.
\ref{Fig_Gap_Model}, which is adapted from Fig. 10.2 on p. 442 in
\cite{Gujrati-book}. The behavior of the entropies and their consequences for
a given cooling rate $r$ have been discussed by us in Ch. 10 in
\cite{Gujrati-book}, and is based on the energy gap model for glasses
discussed there.
%EndExpansion
The form of all these entropies is to ensure that they remain \emph{concave}
functions of the energy. The basic premise of the model is the well-known fact
\cite{Landau} that there exists a \emph{gap} (an energy gap) between the
energies of the perfect crystal and the glass at absolute zero, which is at
least as much as $E_{\text{K}}$; it is equal to $E_{\text{K}}$ for the
\emph{ideal glass}. The gap is due to \emph{defects} that are present in the
glass with respect to the perfect crystal that raise the energy or the
enthalpy of the former \cite{Gujrati-Defects,Gujrati-book}. The additional
energy of a glass $E_{\text{NE}}-E_{\text{K}}$ in a glass, however, is due to
the presence of excitations relative to the ideal glass. It should be observed
that the slope of FG at G is infinite to ensure that the energy of the system
described by FG is $E_{\text{NE}}>E_{\text{K}}$ at absolute zero. The
investigation of the energy gap model in \cite{Gujrati-book} is very relevant
here as it is this non-equilibrium component confinement during the glass
transition that we wish to study in this work. The entropy curve FG in Fig.
\ref{Fig_Gap_Model} denotes the entropy of some arbitrary time-dependent
non-equilibrium state including glasses, and is shown disconnected from the
entropy curve BA0 describing the entropy of the disordered state which
includes the liquid and the supercooled liquid. The latter is a
\emph{stationary }metastable state below the melting temperature $T_{\text{M}%
}$, with the crystal representing the stable state below $T_{\text{M}}$. The
entropy of the ordered state, which results in the crystal below $T_{\text{M}%
}$, is schematically shown by BA0. It is a common practice to call the
supercooled liquid as an equilibrium state when the interest is mainly to
study non-equilibrium states denoted by FG; the terminology is obviously
incorrect if we need to bring the crystal into the discussion. As our interest
in this work is to mainly study time-dependent non-equilibrium states, we will
invariably call the supercooled liquid as the equilibrium liquid, knowing well
its limitation. In a glass transition from the supercooled liquid, the curve
FG emerges out of BACK continuously at some point C, which is determined by
the experimental setup such as the cooling rate $r$.

As the system is not allowed to probe all the disjoint components in time, it
can be investigated by our recently developed non-equilibrium approach in I
and II; see for example, Eqs. (\ref{Heat_Entropy Relation}%
-\ref{Heat_Capacity Integral}). One of the important observations of such
systems is the existence of a residual entropy $S_{\text{R}}$, such as the
entropy at the point G in Fig. \ref{Fig_Gap_Model}. It is the entropy a glass
or a crystal would have at absolute zero, and is related to the possible
number of disjoint components in the phase space
\cite{Palmer,Jackle,Gujrati-Symmetry}. It is merely a measure of the number of
lowest energy microstates of energy $E_{\text{NE}}>E_{\text{K}}$ for a given
\emph{mode of preparation} at absolute zero. The macrostate of the glass or
the crystal is a collection of all these microstates along with their
probability of occurrence for a given mode; see
\cite{Gujrati-Symmetry,GujratiResidualentropy}. The mode of preparation
determines the energies of these microstates, which need not be the lowest
possible energy (such as $E_{\text{K}}$ for the supercooled liquid if we are
concerned with a glass) for the system \cite{Gujrati-book}. The lowest
possible energy microstates represent the equilibrium macrostate at absolute
zero. It should be emphasized that there is no reason for the lowest energy
macrostate to be non-degenerate in principle, although it seems to be true for
a majority of the systems. Therefore, we will assume in this work that
$S_{\text{eq,K}}=0$ for the supercooled liquid at K. This is borne out of
several exact calculations as discussed elsewhere \cite{Gujrati-book}. The
system (glass or crystal) can be in any of lowest energy microstates (and not
the lowest \emph{possible} enrgy microstates) for a given mode of preparation,
and the residual entropy is determined by all these microstates
\cite{Gujrati-Symmetry}.

\subsubsection{Residual Entropy}

The existence of a non-zero residual entropy ($S_{\text{R}}>0$) is very common
in Nature, and does not violate Nernst's postulate, as the latter is
applicable only to equilibrium states with a non-degenerate ground state
\cite[Sect. 64]{Landau}. Its existence was first demonstrated by Pauling and
Tolman \cite{Pauling}; see also Tolman \cite{Tolman}. In addition, the
existence of the residual entropy has been demonstrated rigorously for a very
general spin model by Chow and Wu \cite{Chow}. For ice, the residual entropy
was observed by Giauque and Ashley \cite{Giauque}. Pauling \cite{Pauling-ice}
provided the first numerical estimate for the residual entropy for ice, which
was later improved by Nagle \cite{Nagle}. Nagle's numerical estimate has been
recently verified by simulation \cite{Isakov,Berg}. The numerical simulation
carried out by Bowles and Speedy \cite{Speedy} for glassy dimers also supports
the existence of a residual entropy. Thus, it appears that the support in
favor of the residual entropy is quite strong. We wish to emphasize that what
is customarily called the third law due to Nernst, according to which the
entropy must vanish at absolute zero, is merely a postulate and not a strict
theorem even in equilibrium \cite{Landau,Gujrati-Nernst,Gujrati-Fluctuations}.
Indeed, many exactly solved statistical mechanical models show a non-zero
entropy at absolute zero. Based on these empirical observations, the residual
entropy $S_{\text{R}}$ at G in Fig. \ref{Fig_Gap_Model} has been taken as
non-zero and positive.

We refer to Fig. \ref{Fig_Gap_Model} to emphasize this point once more. As $r$
decreases, G moves not only to the left but also down in the figure. This has
the consequence of lowering $E_{\text{NE}}$ and $S_{\text{R}}$, so that
entropy continues to decrease with temperature to ensure stability of the
glass. The additional energy or enthalpy due to excitation compared to the
ideal glass raises the residual entropy of the glass relative to the ideal
glass:%
\begin{equation}
S_{\text{R}}>S_{\text{eq,K}}=0\text{ for }E_{\text{NE}}>E_{\text{K}}.
\label{Gap_Model_Entropy_Bound}%
\end{equation}

However, it is possible that the residual entropy $S_{\text{R}}$ at G remains
zero for all $r$ as if all glasses are identical at absolute zero for any $r$.
Indeed, the situation shown in Fig. \ref{Fig_Gap_Model} has been challenged
\cite{Hemmen,Thirumalai,Kievelson,Mauro-Gupta,Note1}, because it is argued
that the entropy cannot be estimated correctly in the glass transition region,
the region where component confinement occurs, where \emph{irreversibility}
comes into play. This is certainly true, as can be easily seen form Eqs
(\ref{Entropy_HeatCapacity}) and (\ref{General_Entropy_Calculation}). While
the idea that the irreversibility raises some concern about the inferred
values of the entropy is certainly justified, the main point is not whether
irreversibility is present; see Sect.
\ref{Marker_Irreversibility_Residual_Entropy}. Rather, the question should be:
how much of an error does this create in the inferred values of the entropy?
Thus, the issue at hand is the magnitude of $\Delta S_{\text{B}}(T_{0},P_{0}%
)$. Various estimates support an error that does not exceed more than 5\%.
J\"{a}ckle \cite{Jackle} has also calculated the amount of uncertainty in the
entropy for glycerol due to irreversibility near the glass transition, and
found it to be less than 5\%. Gutzow and Schmelzer \cite{Gutzow} have shown by
citing some earlier analysis that the irreversible contribution to the entropy
in the vicinity of the glass transition is small enough to be practically
neglected. They also provide a very enlightening historical review of the
residual entropy in glasses. Nemilov \cite{Nemilov} comes to a similar
conclusion by suggesting that the error in $S_{\text{R}}$ is no more than
5\%.\ The conclusion is that there is no appreciable error in using
conventional equilibrium thermodynamics by replacing the inequality to an
equality in Eqs. (\ref{Entropy_HeatCapacity_Bounds}-\ref{Entropy_HeatCapacity}%
) in studying glasses, a point also made recently by Goldstein
\cite{Goldstein}. In particular, one can get a highly reliable estimate of the
entropy of a glass (GL) or the supercooled liquid (SCL) without using any
internal order parameter by the following standard integration process using
the measured heat capacity%
\begin{equation}
S(T_{0})=S(0)+%
%TCIMACRO{\tint \limits_{0}^{T_{0}}}%
%BeginExpansion
{\textstyle\int\limits_{0}^{T_{0}}}
%EndExpansion
\frac{C_{P}dT_{0}}{T_{0}}+\Delta S_{\text{C}}. \label{Entropy_HeatCapacity0}%
\end{equation}
More recently, Johari and Khouri \cite{Johari-New} have come to a similar
conclusion by reanalyzing experimental data.

\subsubsection{Entropy Reduction}

It has been suggested \cite{Hemmen,Thirumalai,Kievelson,Mauro-Gupta,Note1}
that as a consequence of the component confinnment, the system is no longer
able to explore \emph{in time} the entire part of the phase space that would
be consistent with the observables (such as the energy) of the system. With
this interpretation in mind, and recognizing the fact that the entropy is
measured by the volume of the accessible phase space, such as that belonging
to a component, it is then argued that the entropy $S$ of the system must
undergo a sudden or rapid \emph{reduction} in its value due to component
confinement \cite{Note1} that is expected to occur at the glass transition,
but the experiments are unable to capture this because of the irreversibility.
We will call this entropy reduction scenario the \emph{unconventional view}
(UV) of the glass transition in this work. In this scenario, either FG in Fig.
\ref{Fig_Gap_Model} remains disconnected from BA0 to account for a sudden
discontinuous reduction or touches BA0 at some point C but falls off rapidly
from C to give rise to a rapid but continuous drop. In the latter scenario,
the portion of FG to the immediate left of C will have to change its curvature
so that it will be no longer concave, a point of critical importance that will
be discussed at several occasions in this work. It is impossible to
continuously join two disjoint concave functions without creating a loss of concavity.

If it happens that the entropy $\widetilde{S}$ of the medium concurrently
undergoes a sudden or rapid \emph{rise} by an amount such that it not only
compensates the reduction in $S$, but add some more so that $S_{0}$ does not
decrease, then the rapid reduction in $S$ would not violate the second law for
the isolated system as given in Eq. (\ref{Second_Law}). However, in this case
also, the irreversible entropy change in the system must still satisfy the
second law: $d_{\text{i}}S(t)\geq0$. Therefore, our third aim is to
investigate if such a scenario can be supported by the second law.

Our last goal is to use the second law to investigate the idea of the residual
entropy from a statistical point of view and to prove that the situation in
Fig. \ref{Fig_Gap_Model} with a non-zero residual entropy is borne out of
experiments. This will be in contradiction with the idea of an entropy
reduction in the unconventional view. The enthalpy $H$ and the volume $V$ are
known to be continuous during the glass transition
\cite{Gutzow-Book,GoldsteinSimha,Jones,Nemilov-Book,Gujrati-book}. Thus, any
entropy reduction will leave its imprint on the behavior of the Gibbs free
energy. Thus, we will also investigate the behavior of the Gibbs free energy
in this work.

The other relevant issue, which to the best of our knowledge has not been
discussed in the literature so far for unexplained reasons, is whether the
experimental data support a non-zero residual entropy even if the
irreversibility may be quite appreciable, as would happen in a rapid quench.
From above, it is clear that the experimental determination of the residual
entropy $S_{\text{expt}}(0)$ invariably yields a non-zero value. We derive
later a very simple consequence of the glassy irreversibility in the form of
an inequality; see Eq. (\ref{General_Entropy_Calculation}). The inequality
clearly supports a non-zero residual entropy, even if the irreversible
contribution is appreciable; see Eq. (\ref{Residual_Entropy_Bound}). It just
happens that in most cases, the irreversible contribution turns out to be
minimal. However, the reality of the residual entropy is based not on the
minimal irreversible contribution; it is rather based on the inequality in Eq.
(\ref{General_Entropy_Calculation}). Thus, the \emph{entropy reduction} due to
\emph{confinement} in the unconventional view cannot be attributed simply to
the presence of irreversibility. Its root-cuse, if it exists, has to be found elsewhere.

\subsection{Goal and Outline}

Our aim in this work is to follow the recently developed non-equilibrium
thermodynamics \cite{Guj-NE-I,Guj-NE-II,Gujrati-Symmetry} to specifically
derive non-equilibrium thermodynamic principles, which should reduce to the
well-known thermodynamic principles for equilibrium bodies
\cite{Landau,Huang,Tolman}. We must ensure that these principles do not
violate the second law in anyway. We start by assuming the existence and
continuity of $S_{0}$, and follow I and II. The existence assumption should
not be surprising as the existence of $S_{0}$ is required to state the second
law for an isolated system $\Sigma_{0}$; see Eq. (\ref{Second_Law}). The
equality in Eq. (\ref{Second_Law}) occurs only when the isolated system is in
equilibrium. We will prove here that the existence and continuity of $S_{0}$
implies the existence and \emph{continuity} of the Gibbs free energy
\cite{Guj-NE-I,Guj-NE-II}
\begin{equation}
G(T_{0},P_{0},t)=E(t)-T_{0}S(t)+P_{0}V(t) \label{Gibbs_Free_Energy}%
\end{equation}
for the open system~$\Sigma$ at fixed temperature $T_{0}$ and pressure $P_{0}$
of the medium. (Even though it is not a common terminology, we will use "open"
to denote any body that is not isolated in this work.) The continuity of
$S_{0}$ is easily traced to the continuity of microstate probabilities, as
discussed in Sect. \ref{Marker_NonEq-S}. The second law for the open system
reduces to the well-known inequality
\begin{equation}
\frac{dG(t)}{dt}\leq0; \label{Gibbs_Free_Energy_Variation}%
\end{equation}
the equality occurs when the system achieves equilibrium with the medium; this
is exemplified by the curve ACK in Fig. \ref{Fig_Gap_Model}. It is evident
from Fig. \ref{Fig_Gibbs_Free_Energy} that glass free energies
$G_{\text{GL,CV}}(T_{0})$ and $G_{\text{GL,UV}}(T_{0})$ approach
$G_{\text{SCL}}(T_{0})$\ from above, in conformity with Eq.
(\ref{Gibbs_Free_Energy_Variation}). Thus, their behavior during relaxation is
not sufficient to rule out either one as unphysical. On the other hand, the
continuity of the Gibbs free energy immediately rules out any discontinuous
entropy reduction due to component confinement as discussed above but does not
rule out a continuous entropy reduction.

The continuity of the Gibbs free energy also establishes the \emph{continuity}
of the entropy of an open system in internal equilibrium by recognizing that
the enthalpy of the system remains continuous at the glass transition. The
existence of internal equilibrium still allows irreversible relaxation in the
system. We find that as a consequence of the second law, the entropy of the
system in internal equilibrium decreases during relaxation, thus validating
Eq. (\ref{S_CV_Variation}) and invalidating Eq. (\ref{S_ELV_Variation}). The
latter requires the entropy to increase during relaxation as shown by the
relaxation arrow for $S_{\text{GL,UV}}$ in Fig. \ref{Fig_EntropyLoss}. Thus,
the second law supports a continuous entropy $S$ during component confinement.
On the other hand, the Gibbs free energies of the two glasses satisfy the
second law in Eq. (\ref{Gibbs_Free_Energy_Variation}). In view of the
incorrect behavior of $S_{\text{GL,UV}}$, this is a very surprising result. It
appears that $S_{\text{GL,UV}}(T_{0})$ and $G_{\text{GL,UV}}(T_{0})$ do not
have the same content. We will find an explanation of this surprising
statement later.

The statement about the continuity of the entropy of the system can be made
stronger by removing the requirement of internal equilibrium by obtaining a
statistical formulation for it. However, we only need the above weaker
continuity statement to rule out the unconventional view of the entropy during
the glass transition and to support Fig. \ref{Fig_Gap_Model}. The statistical
formulation of the entropy of an isolated system in terms of its microstates
is given by Eq. (\ref{Conventional_Entropy}) in compliance with the second law
and which satisfies the above thermodynamic principles. We then discuss how
the entropy is defined for an open system under all conditions in terms of
microstate probabilities. This derivation also establishes the
\emph{existence} and \emph{continuity} of the entropy of an open system by
recognizing that the entropy of the medium remains continuous during
relaxation. This again establishes that continuity of entropy when component
confinement occurs. The continuity of the entropy of a body is a generic
feature of a statistical system, and is formulated as one of the thermodynamic
principles. Therefore, it should also not come as a surprise that the
assumption of the existence of the entropy for non-equilibrium states also
forms the cornerstone of the conventional non-equilibrium thermodynamics
\cite{Donder,deGroot,Prigogine}. We then proceed to explain the residual
entropy in statistical terms and show that there is no reason for it to
vanish. We also show that irreversibility alone cannot justify any entropy reduction.

The layout of the paper is as follows. We briefly discuss the two alternative
views of the glass transition in Sect. \ref{Marker_Two_Views}, which is then
followed by a short review of some of the useful concepts and results derived
earlier \cite{Guj-NE-I,Guj-NE-II,Gujrati-Symmetry} in Sect.
\ref{Marker_Useful_Results}. Non-equilibrium statistical entropy is introduced
in Sect. \ref{Marker_NonEq-S} and the concept of residual entropy is
discussed. We then list and discuss important thermodynamic principles that
should be valid for a body in internal equilibrium in Sect.
\ref{Sect_Thermodynamic_Principles}. Sect. \ref{Marker_Stability} is the
important section, where we discuss the consequences of the non-equilibrium
thermodynamics. We discover in Sect. \ref{Marker_Inconsistency} that the
unconventional view is internally inconsistency. The last section contains our conclusions.

\section{Two Views of Glasses\label{Marker_Two_Views}}

\subsection{Conventional Glass View (CV)}

\begin{figure}
[ptb]
\begin{center}
\includegraphics[
trim=0.000000in 0.400920in 0.160680in 0.000000in,
height=3.0398in,
width=5.2416in
]%
{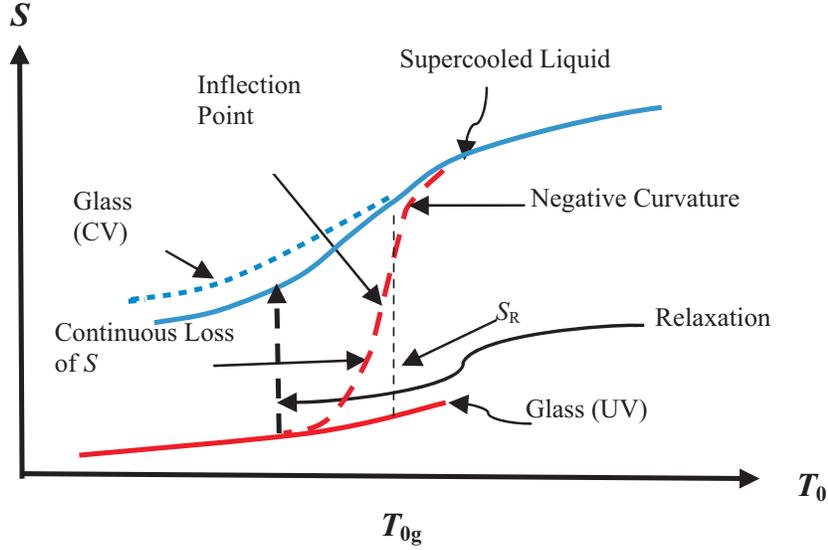}%
\caption{The entropy of the supercooled liquid ($S_{\text{SCL}}$: upper solid
curve) and of the CV-glass ($S_{\text{GL,CV}}$: short dashed curve) in the
conventional view (CV) are shown schematically. We are considering an isobaric
cooling experiment. The entropy $S_{\text{GL,CV}}$ of the CV-glass
continuously emerges out of $S_{\text{SCL}}$\ at $T_{0\text{g}}$\ and
approaches the residual entropy at absolute zero (not shown). Schematic form
of the entropy $S_{\text{GL,UV}}$ in the unconventional view (UV) is shown by
a discontinuous entropy loss (thin dashed vertical line of magnitude
$S_{\text{R}}$) at $T_{0\text{g}}$\ or a continuous entropy loss (dashed
curve) of $S_{\text{R}}$ from $S_{\text{SCL}}$\ to the lower solid
curve\ during the glass transition across $T_{0\text{g}}$. Note the presence
of an inflection point in the dashed piece of $S_{\text{GL,UV}}$; the latter
approaches $0$ at absolute zero upon extrapolation (not shown). In the
presence of an additional variable $\xi$, see text, the entropy curves turn
into surfaces defined over the $T_{0}-\xi$ plane. }%
\label{Fig_EntropyLoss}%
\end{center}
\end{figure}

The small amount of irreversible contribution to the entropy of glasses does
not mean that glasses are in equilibrium (with the surrounding medium);
rather, it implies that the fast degrees of freedom in glasses have
equilibrated, and the slow degrees of freedom, although not yet equilibrated,
are changing so slowly as to be almost unchanged over the observation time.
More details about this point can be found in I. For most glasses, the success
of Eq. (\ref{Entropy_HeatCapacity0}) rests on the premise that within $5\%$ of
the residual entropy, as noted above, $d_{\text{i}}S\simeq0$ so that $dS\simeq
d_{\text{e}}S.$ As a consequence, $S(0)$ for a glass in Eq.
(\ref{Entropy_HeatCapacity0}) is correct within $5\%$; its entire magnitude is
only weakly affected by the irreversibility encountered during the glass
transition. This view will be called the \emph{conventional view} (CV) of the
glass transition. Gutzow and Schmelzer \cite{Gutzow} draw an analogy of
$d_{\text{i}}S\simeq0$ in slowly varying glasses with Prigogine's principle of
minimum entropy production \cite{Prigogine}. However, we will not assume
$d_{\text{i}}S\simeq0$ so as to make our argument as general as possible.
Later in this work, we will derive an almost identical expression to Eq.
(\ref{Entropy_HeatCapacity0}) as an identity; see\ Eq.
(\ref{Entropy_HeatCapacity_Identity}). In the conventional view, the entropy
of the glass is schematically represented by the short dashed curve in Fig.
\ref{Fig_EntropyLoss}; the upper solid curve represents the entropy of the
supercooled liquid (SCL) that would be eventually observed if we wait long
enough for the system to equilibrate. The point where $S_{\text{GL,CV}}%
(T_{0)})$ branches out of $S_{\text{SCL}}(T_{0})$ is the glass transition
temperature $T_{0\text{g}}$ and is determined by the choice of observation
time $\tau_{\text{obs}}$. As the entropy of the conventional glass lies above
the supercooled liquid entropy%
\[
S_{\text{GL,CV}}(T_{0},t)>S_{\text{SCL}}(T_{0})\text{ \ \ for \ \ }%
T_{0}<T_{0\text{g}},
\]
the entropy of the glass actually decreases in time%
\begin{equation}
\frac{dS_{\text{GL,CV}}(T_{0},t)}{dt}<0. \label{S_CV_Variation}%
\end{equation}
The entropy of the conventional glass (blue short dashed curve) approaches
$S_{\text{R}}>0$ if we extrapolate to absolute zero for most glasses. In this
view, the glass continuously emerges from the supercooled liquid at
$T_{0\text{g}}$. Both entropy functions are continuous with only gradual
variation determined by their respective heat capacities.%
%TCIMACRO{\FRAME{ftbpFU}{5.2416in}{3.0398in}{0pt}{\Qcb{The entropy of the
%supercooled liquid ($S_{\text{SCL}}$: upper solid curve) and of the CV-glass
%($S_{\text{GL,CV}}$: short dashed curve) in the conventional view (CV) are
%shown schematically. We are considering an isobaric cooling experiment. The
%entropy $S_{\text{GL,CV}}$ of the CV-glass continuously emerges out of
%$S_{\text{SCL}}$\ at $T_{0\text{g}}$\ and approaches the residual entropy at
%absolute zero (not shown). Schematic form of the entropy $S_{\text{GL,UV}}$ in
%the unconventional view (UV) is shown by a discontinuous entropy loss (thin
%dashed vertical line of magnitude $S_{\text{R}}$) at $T_{0\text{g}}$\ or a
%continuous entropy loss (dashed curve) of $S_{\text{R}}$ from $S_{\text{SCL}}%
%$\ to the lower solid curve\ during the glass transition across $T_{0\text{g}%
%}$. Note the presence of an inflection point in the dashed piece of
%$S_{\text{GL,UV}}$; the latter approaches $0$ at absolute zero upon
%extrapolation (not shown). In the presence of an additional variable $\xi$,
%see text, the entropy curves turn into surfaces defined over the $T_{0}-\xi$
%plane. }}{\Qlb{Fig_EntropyLoss}}{figentropyloss2_2003.eps}%
%{\special{ language "Scientific Word";  type "GRAPHIC";
%maintain-aspect-ratio TRUE;  display "USEDEF";  valid_file "F";
%width 5.2416in;  height 3.0398in;  depth 0pt;  original-width 4.331in;
%original-height 2.8115in;  cropleft "0";  croptop "1";  cropright "0.9629";
%cropbottom "0.1426";
%filename 'FigEntropyLoss2_2003.eps';file-properties "XNPEU";}}}%
%BeginExpansion
%EndExpansion

\subsection{Unconventional Glass View (UV)}

Despite the success of the conventional view that provides a justification of
the residual entropy, there have been several attempts to argue that the
residual entropy is not a physical concept
\cite{Hemmen,Thirumalai,Kievelson,Mauro-Gupta,Note1}. It has been suggested
that the entropy of a glass should approach $0$ at absolute zero, regardless
of how it is prepared. In other words, the entropy of a glass has no memory of
its mode of preparation at absolute zero, notwithstanding the fact that most
glasses do not represent equilibrium states. The total loss of memory of the
mode of preparation is the hallmark property of an equilibrium state, while a
non-equilibrium state remains strongly correlated with its mode of preparation
until it fully equilibrates, which is reflected in its relaxation towards
equilibrium. Despite this, the above suggestion has been made repeatedly in
the literature. If it is true, it will require either an \emph{abrupt} or a
\emph{gradual }but \emph{rapid reduction} in the entropy as the supercooled
liquid (SCL) turns into a glass (GL) at (in the former case) or around (in the
latter case) $T_{0\text{g}}$ as we cool SCL. The amount by which the entropy
must decrease during the glass transition (by confinement to one of the
possible exponentially large number of components) should be comparable to
$S_{\text{R}}$. This is the unconventional view (UV) described above. The
abrupt entropy reduction during the glass transition is shown by the thin
vertical dashed line of height $S_{\text{R}}$ at $T_{0\text{g}}$ to the lower
solid entropy curve of the unconventional glass shown as Glass (UV) in Fig.
\ref{Fig_EntropyLoss}. The gradual entropy reduction is shown by the dashed
red portion with an inflection point, which falls very rapidly below the
entropy of the supercooled liquid. The total additional entropy reduction is
equal to $S_{\text{R}}$. After the drop, the dashed curve connects with the
schematic lower solid Glass (UV), which when extrapolated to absolute zero
will now give a vanishing entropy, thus resulting in no residual entropy. We
note that because of the entropy reduction,%
\[
S_{\text{GL,UV}}(T_{0},t)<S_{\text{SCL}}(T_{0})\text{ \ \ for \ \ }%
T_{0}<T_{0\text{g}},
\]
so that the entropy of the glass actually increases in time%
\begin{equation}
\frac{dS_{\text{GL,UV}}(T_{0},t)}{dt}>0, \label{S_ELV_Variation}%
\end{equation}
which is of opposite nature to that in Eq. (\ref{S_CV_Variation}).%
%TCIMACRO{\FRAME{ftbpFU}{4.7037in}{3.9963in}{0pt}{\Qcb{Schematic form of the
%free energy $G_{\text{SCL}}$(lower solid curve), $G_{\text{GL,CV}}$ (short
%dashed curve) and $G_{\text{GL,UV}}$ (vertical jump at $T_{0\text{g}}$ or a
%continuous dashed piece along with the solid upper curve). The dashed piece
%has an inflection point A, so that it has a region of positive curvature at
%higher temperatures.There is no inflection point in $G_{\text{GL,CV}}$ and
%$G_{\text{SCL}}$.\ In the presence of an additional variable $\xi$, see text,
%the Gibbs free energy curves turn into surfaces defined over the $T_{0}-\xi$
%plane. }}{\Qlb{Fig_Gibbs_Free_Energy}}{gibbs_free_energy_elv.eps}%
%{\special{ language "Scientific Word";  type "GRAPHIC";
%maintain-aspect-ratio TRUE;  display "USEDEF";  valid_file "F";
%width 4.7037in;  height 3.9963in;  depth 0pt;  original-width 5.1949in;
%original-height 5.1076in;  cropleft "0";  croptop "0.9609";  cropright "1";
%cropbottom "0.0976";
%filename 'gibbs_free_energy_elv.eps';file-properties "XNPEU";}}}%
%BeginExpansion
\begin{figure}
[ptb]
\begin{center}
\includegraphics[
trim=0.000000in 0.498502in 0.000000in 0.199707in,
height=3.9963in,
width=4.7037in
]%
{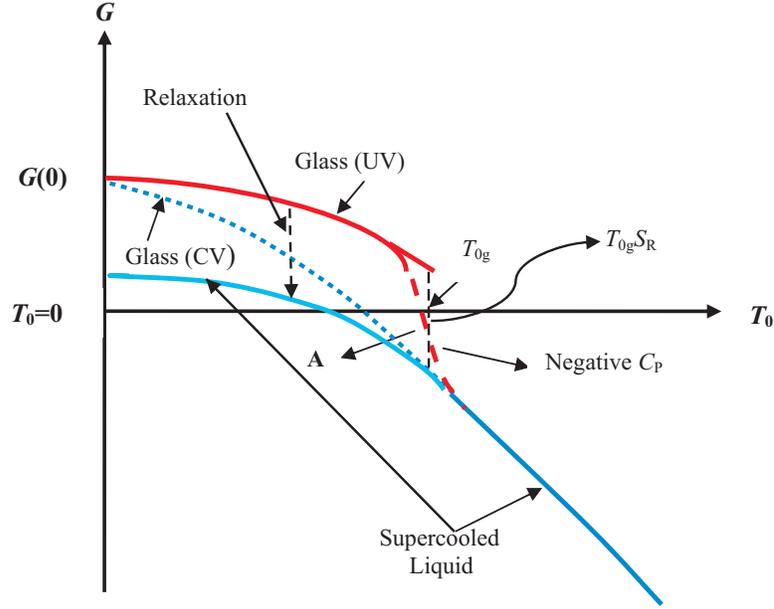}%
\caption{Schematic form of the free energy $G_{\text{SCL}}$(lower solid
curve), $G_{\text{GL,CV}}$ (short dashed curve) and $G_{\text{GL,UV}}$
(vertical jump at $T_{0\text{g}}$ or a continuous dashed piece along with the
solid upper curve). The dashed piece has an inflection point A, so that it has
a region of positive curvature at higher temperatures.There is no inflection
point in $G_{\text{GL,CV}}$ and $G_{\text{SCL}}$.\ In the presence of an
additional variable $\xi$, see text, the Gibbs free energy curves turn into
surfaces defined over the $T_{0}-\xi$ plane. }%
\label{Fig_Gibbs_Free_Energy}%
\end{center}
\end{figure}
%EndExpansion

\subsection{Gibbs free energy:\ VConventional versus Unconventional}

Thus, it appears that there are two competing views at present as far as the
entropy of the glass is concerned with contradicting behavior of the entropy
during relaxation as seen from Eqs. (\ref{S_CV_Variation}%
-\ref{S_ELV_Variation}). However, they both accept the same entropy function
$S_{\text{SCL}}(T_{0})$ of the supercooled liquid before it turn into a glass,
and is given by the solid blue curve indicated by "Supercooled Liquid." As the
enthalpy $H$ and the volume $V$ of the glass are known to be \emph{continuous}
during the glass transition
\cite{GoldsteinSimha,Jones,Nemilov-Book,Gutzow-Book}, both views agree on the
same functions $H_{\text{GL}}(T_{0})$ and $V_{\text{GL}}(T_{0})$\ for the
glass. The Gibbs free energy $G(T_{0})$ has been identified in I as
$G(T_{0})\equiv H(T_{0})-T_{0}S(T_{0})$ in terms of the entropy of the
system$.$ Conversely, the entropy can be obtained from the Gibbs free energy.
Then the continuity of the entropy in the conventional view ensures that the
Gibbs free energy $G_{\text{GL,CV}}(T_{0})\equiv H_{\text{GL}}(T_{0}%
)-T_{0}S_{\text{GL,CV}}(T_{0})$ of the glass is also continuous. This is shown
by the short dashed piece in Fig. \ref{Fig_Gibbs_Free_Energy} that emerges out
of the lower solid curve $G_{\text{SCL}}(T_{0})$ without any inflection point.
In the unconventional view, the Gibbs free energy abruptly jumps upwards by an
amount close to $T_{\text{g}}S_{\text{R}}$ for an abrupt reduction of entropy
or rises rapidly along the dashed curve with an inflection point to meet
continuously with the upper solid curve to give $G_{\text{GL,UV}}(T_{0})\equiv
H_{\text{GL}}(T_{0})-T_{0}S_{\text{GL,UV}}(T_{0})$ in Fig.
\ref{Fig_Gibbs_Free_Energy}. Both Gibbs free energies coincide at absolute
zero:%
\begin{equation}
G(0)\equiv G_{\text{GL,CV}}(0)\equiv G_{\text{GL,UV}}(0)=H_{\text{GL}}(0),
\label{Common_Gibbs_Free_Energy_at_0}%
\end{equation}
as shown in Fig. \ref{Fig_Gibbs_Free_Energy}. Otherwise,%
\[
G_{\text{GL,CV}}(T_{0}>0)<G_{\text{GL,UV}}(T_{0}>0)\ \ \ \ \text{for
\ \ }T_{0}<T_{0\text{g}}.
\]

The abrupt entropy reduction in the unconventional view has been recently
criticized by Goldstein \cite{Goldstein}, Gutzow and Schmelzer \cite{Gutzow},
Nemilov \cite{Nemilov}, Gujrati
\cite{GujratiResidualentropy,Gujrati-Symmetry,Gujrati-Comments}, and Johari
\cite{Johari}. We will not go into their arguments here for which the reader
is advised to go to the original source; however, Johari \cite{Johari}
provides a good summary of these approaches. In addition, Kozliak and Lambert
\cite{Kozliak}\ provide a new look at the issue of the residual entropy. These
attempts are different from the approach we take here, although we arrive at a
similar conclusion. These previous attempts neither consider a continuous
entropy reduction nor do they visit the issue of how to define the entropy for
non-equilibrium states from first principles, or how to interpret the residual
entropy from a statistical point of view, the issues that guide us here.
Therefore, the previous attempts do not reveal a pathological aspect of the
continuous entropy reduction in that the behavior of the entropy
$S_{\text{GL,UV}}(T_{0})$ in Fig. \ref{Fig_EntropyLoss} is not consistent with
the entropy obtained from the Gibbs free energy $G_{\text{GL,UV}}(T_{0})$ in
Fig. \ref{Fig_Gibbs_Free_Energy}. This \emph{internal inconsistency} poses a
major problem for the unconventional view.

If the entropy can be defined for a non-equilibrium body from first principles
and if it satisfies the second law, then we should be able to test whether
component confinement results in an entropy reduction or whether the residual
entropy is real for a non-equilibrium body.

\section{Important Concepts and Useful Results from I \&
II\label{Marker_Useful_Results}}

Quantities pertaining to $\Sigma_{0}$ are denoted by a subscript $0$, for
$\Sigma$ without any subscript and for $\widetilde{\Sigma}$ by a tilda at the
top. The system, although macroscopically large, is considered to be a very
small part compared to the medium $\widetilde{\Sigma}$, so that the system can
only create a very weak disturbance on the medium. As said earlier, a body
refers to any of the three systems $\Sigma$, $\widetilde{\Sigma}$ and
$\Sigma_{0}$. For the sake of convenience, quantities pertaining to a body
will be denoted without any subscript.

\subsection{State variables and internal
equilibrium\label{Marker_Internal Variables}}

We first need to clarify the concept of the macroscopic state of a body. As
equilibrium and non-equilibrium thermodynamics is an experimental science, it
must be based on observables, which we take to be extensive and collectively
denote by $\mathbf{X}(t)$, $t$\ being the time measured from some convenient
reference point. Any thermodynamic state function must be a function of state
variables $\mathbf{Z}(t)$ consisting of observable $\mathbf{X}(t)$\ and
internal variable $\mathbf{I}(t)$. The observables are independent
\emph{continuous }variables that can be controlled and manipulated by an
observer so that their values will allow the observer to differentiate between
different macroscopic states\ (macrostates) of the same system. Of the
observables, the energy $E$, volume $V$ and the number of particles $N$ play
an important role. The observables are important as they remain fixed for an
isolated system. The internal variables (see II for details, where they are
denoted collectively by $\mathbf{I}(t)\boldsymbol{\equiv}\{I_{k}(t)\}$) cannot
be controlled by the observer, but play a very important role during
relaxation in non-equilibrium bodies such as glasses. In addition, the state
function may also be an explicit function of time $t$. It is useful to replace
$\mathbf{I}(t)$ by the \emph{dimensionless} vector $\boldsymbol{\xi}(t)$
$\boldsymbol{\equiv}\{\xi_{k}(t)\}$ defined according to%
\[
\xi_{k}(t)\boldsymbol{\equiv}\frac{I_{k}(t)-I_{k}(\infty)}{I_{k}%
(0)-I_{k}(\infty)}.
\]
The variables $\xi_{k}$ are usually known as "internal order parameters" or
the "degree of advancement." Being a state function, the entropy of a body,
whether in equilibrium or not, is a \emph{continuous} function of the state
variables and will be expressed as $S(\mathbf{Z}(t),t)$ or $S(\mathbf{X}%
(t),\mathbf{\xi}(t)\boldsymbol{,}t)$.

When the body refers to the isolated system $\Sigma_{0}$, we should remember
that its observables are fixed so that $\mathbf{X}_{0}$ has no $t$-dependence.
The entropy of $\Sigma_{0}$ must be expressed as $S_{0}(\mathbf{X}%
_{0}\boldsymbol{,}\mathbf{I}_{0}(t),t)$ or $S_{0}(\mathbf{X}_{0}%
\boldsymbol{,}\boldsymbol{\xi}_{0}(t),t)$ while it is relaxing towards its
equilibrium state. Once it reaches equilibrium, its entropy can only be
described as a function of its observables, so that it must be expressed as
$S_{0\text{,eq}}(\mathbf{X}_{0})$, which is now a constant, independent of
$t$. The explicit $t$-dependence has disappeared. At the same time,
$\boldsymbol{\xi}_{0}(t)$ is also no longer $t$-dependent. This means that in
equilibrium, $\boldsymbol{\xi}_{0\text{,eq}}$ must become constant, which from
its definition above must \emph{vanish} identically:%
\begin{equation}
\boldsymbol{\xi}_{0\text{,eq}}\equiv0. \label{Eq_order-parameter}%
\end{equation}
Moreover, as discussed in II, $\mathbf{I}_{0}(t)$ is no longer an independent
variable; it becomes a function of the observable $\mathbf{X}_{0}$:%
\[
\mathbf{I}_{0\text{,eq}}=\mathbf{I}_{0\text{,eq}}(\mathbf{X}_{0}).
\]
For an open system in equilibrium, $\boldsymbol{\xi}_{\text{eq}}$ also must
become constant in equilibrium; similarly, $\mathbf{I}_{\text{eq}}$ is also no
longer an independent variable; it becomes a function of the observable
$\mathbf{X}_{\text{eq}}$:%
\[
\boldsymbol{\xi}_{\text{eq}}\equiv0,\ \ \mathbf{I}_{\text{eq}}=\mathbf{I}%
_{\text{eq}}(\mathbf{X}_{\text{eq}}).
\]

The case when the entropy of a body has no explicit $t$-dependence, so that
the entropy appears as
\begin{equation}
S(\mathbf{X}(t),\boldsymbol{\xi}(t)), \label{internal_equilibrium}%
\end{equation}
is quite special as discussed in I and II. In this case, the system is said to
be in \emph{internal equilibrium}. Accordingly,
\begin{equation}
\frac{\partial S}{\partial t}=0 \label{Eq_Conds_t}%
\end{equation}
under internal equilibrium. The entropy has the maximum possible value for
given $\mathbf{Z}(t)$. In this case, the derivatives%
\begin{equation}
\mathbf{y}(t)\mathbf{\equiv}\frac{\mathbf{Y}(t)}{T(t)}\equiv\frac{\partial
S(\mathbf{Z}(t))}{\partial\mathbf{X}(t)},\ \mathbf{a}(t)\mathbf{\equiv}%
\frac{\mathbf{A}(t)}{T(t)}\equiv\frac{\partial S(\mathbf{Z}(t))}%
{\partial\mathbf{I}(t)} \label{Field_Affinity}%
\end{equation}
of the entropy with respect to $\mathbf{Z}(t)$ give the field $\mathbf{Y}(t)$
and affinity $\mathbf{A}(t)$ characterizing the body; see II and below. They
are defined only when the body is in internal equilibrium or in equilibrium.
Otherwise, the derivatives do not have any physical significance.

It will prove useful to think, at least in a formal sense, that the internal
variable can also be kept fixed. Consider the system in internal equilibrium
at the instance when its state variables are
\[
\mathbf{X}_{\text{IE}}\mathbf{\equiv X}(t),\ \mathbf{I}_{\text{IE}%
}\mathbf{\equiv I}(t),
\]
and its field and affinity vectors are
\[
\mathbf{Y}_{\text{IE}}\mathbf{\equiv Y}(t),\ \mathbf{A}_{\text{IE}%
}\mathbf{\equiv A}(t).
\]
The medium $\widetilde{\Sigma}$ is characterized by $\mathbf{Y}_{0}$, and
$\mathbf{A}_{0}=0$, and will be expressed as $\widetilde{\Sigma}$%
($\mathbf{Y}_{0},0)$; see II. We now disconnect this system from the medium
and bring it in contact with another medium $\widetilde{\Sigma}$%
($\mathbf{Y}_{\text{IE}},\mathbf{A}_{\text{IE}})$\ that is characterized by
$\mathbf{Y}_{\text{IE}}$ and $\mathbf{A}_{\text{IE}}$. The system is in
equilibrium with this medium, with the average values of its observables and
internal variables given by $\mathbf{X}_{\text{IE}}$ and $\mathbf{I}%
_{\text{IE}}$, respectively. Being in equilibrium, it must satisfy all the
requirements of stability. In particular, the heat capacity of the system must
be non-negative:
\begin{equation}
C_{\mathbf{Y}_{\text{IE}}^{\prime},\mathbf{A}_{\text{IE}}}\geq0,
\label{Heat_Capacity_Stability}%
\end{equation}
where $\mathbf{Y}_{\text{IE}}^{\prime}$ represents all fields except
$T_{\text{IE}}$. One can also formally consider this equilibrium system as an
isolated system with fixed $\mathbf{X}_{\text{IE}}$ and $\mathbf{I}%
_{\text{IE}}$ by disconnecting it from the second medium. This isolated system
is still in equilibrium, as its entropy is at its maximum for given
$\mathbf{X}_{\text{IE}}$ and $\mathbf{I}_{\text{IE}}$. The heat capacity of
such a system should also be non-negative%
\begin{equation}
C_{\mathbf{X}_{\text{IE}}^{\prime},\mathbf{I}_{\text{IE}}}\geq0,
\label{Heat_Capacity_Stability_0}%
\end{equation}
where $\mathbf{X}_{\text{IE}}^{\prime}$ denotes all observables except energy
$E(t)$. If we connect this system to a medium which only controls
$E_{\text{IS}}$ and $V_{\text{IS}}$ of the system by its fixed $T_{\text{IE }%
}$and $P_{\text{IE}}$, respectively, with the remaining observables
$\mathbf{X}_{\text{IE}}^{^{\prime\prime}}$ and $\mathbf{I}_{\text{IE}}$ kept
fixed, then the stability of such a system ensures that the following heat
capacity is also non-negative:%
\begin{equation}
C_{P_{\text{IE}},\mathbf{X}_{\text{IE}}^{^{\prime\prime}},\mathbf{I}%
_{\text{IE}}}\geq0. \label{Heat_Capacity_Stability_1}%
\end{equation}

\subsection{Gibbs free energy of the system\label{Marker_Gibbs_Free_Energy}}

From now on, we will considerably simplify our discussion by considering only
$E,V$ and $N$ as the only observables of a body, and allow only one internal
variable in describing it, which we denote simply by $\xi$. Moreover, we will
keep $N$ fixed and allow the possibility of fluctuating $E$ and $V$ due to
exchange with another body, the medium. Accordingly, we will not exhibit $N$
in expressing any state function of the body. Thus, $\mathbf{X}$\ will refer
to $E,V$ and $\mathbf{Z}$\ will refer to $E,V,\xi$ from now on. We also assume
that the medium is in internal equilibrium, so that its entropy $\widetilde
{S}(\widetilde{\mathbf{Z}})$ does not have an explicit $t$-dependence (so that
Eq. (\ref{Eq_Conds_t}) is satisfied), although it is not in equilibrium with
the system. It follows from the internal equilibrium of $\widetilde{\Sigma}$
that the associated fields are%
\begin{equation}
\frac{1}{T_{0}}\equiv\left.  \left(  \frac{\partial\widetilde{S}}%
{\partial\widetilde{E}}\right)  \right\vert _{\mathbf{Z}_{0}},\ \ \frac{P_{0}%
}{T_{0}}\equiv\left.  \left(  \frac{\partial\widetilde{S}}{\partial
\widetilde{V}}\right)  \right\vert _{\mathbf{Z}_{0}}.
\label{Temp_Pressure_Medium}%
\end{equation}
Here, $\mathbf{Z}_{0}$ denotes $E_{0},V_{0}$ and $\xi_{0}$ of $\Sigma_{0}$.
The affinity $A_{0\xi}=0$ for the medium. The entropy $S_{0}(t)$ of
$\Sigma_{0}$ is a continuous function of time $t$, at least
twice-differentiable with respect to its extensive variables, and satisfies
Eq. (\ref{Second_Law}). As discussed in I and II, it can be written as the sum
of the entropies $S(t)$ of the system and $\widetilde{S}(t)$ of the medium%
\begin{equation}
S_{0}(E_{0},V_{0},\xi_{0},t)=S(E,V,\xi,t)+\widetilde{S}(\widetilde
{E},\widetilde{V},\widetilde{\xi}) \label{Total_Entropy}%
\end{equation}
by invoking their quasi-independence. We will provide a direct proof of this
sum later in Sect. \ref{Marker_NonEq-S} when we identify the entropies as
statistical quantities in terms of probabilities of microstates.\textbf{ }In
terms of
\begin{equation}
H(t)\equiv E(t)+P_{0}V(t),\ G(t)\equiv H(t)-T_{0}S(t)=E(t)-T_{0}%
S(t)+P_{0}V(t), \label{Free_Energies}%
\end{equation}
which are the time-dependent Gibbs free energy and the enthalpy, respectively,
of the system $\Sigma$ at fixed $T_{0},P_{0}$ and $A_{0\xi}=0$, we have
\begin{equation}
S_{0}(t)-\widetilde{S}_{0}=S(t)-H(t)/T_{0}=-G(t)/T_{0},
\label{Gibbs_Free_Energy_Entropy_Relation}%
\end{equation}
as shown in I. where $\widetilde{S}_{0}\equiv\widetilde{S}(E_{0},V_{0},\xi
_{0})$ is independent of the system.

\begin{theorem}
The Gibbs free energy $G(T_{0},P_{0},t)$ of the system given in Eq.
(\ref{Gibbs_Free_Energy}) is determined by the temperature and pressure of the
medium and is equal to
\begin{equation}
G(T_{0},P_{0},t)\equiv T_{0}[\widetilde{S}_{0}-S_{0}(t)].
\label{Gibbs_Free_Energy_Entropy_Relation_0}%
\end{equation}

\end{theorem}

This is an important observation for the following two reasons. Firstly, it
shows that the Gibbs free energy is defined even though we may not define the
temperature and pressure of the system, a situation that occurs when the
system is not even in internal equilibrium. Of course, we are assuming that
the energy, volume and entropy of the system are defined under all
circumstances. While there is less doubt that this is possible for the
mechanical quantities (such as energy and volume), the concept of entropy
valid under all conditions requires some care; see I and Sect.
\ref{Marker_NonEq-S} later\textbf{.} In engineering context, the above Gibbs
free energy $G(T_{0},P_{0},t)$ is also known as \emph{exergy} or
\emph{availability }\cite{Keenan}.

Secondly, the above conclusion makes an important statement about the
\emph{continuity} of $G(t)$, which we now explain. We first recall that the
entropy $S_{0}(t)$ is a continuous function. Now, $\widetilde{S}_{0}$ is the
entropy of the medium in internal equilibrium (but at $E_{0},V_{0},\xi_{0}$).
\ As said above, the properties of a body in internal equilibrium is similar
to those of the same body in equilibrium, when treated as an isolated system.
Since the entropy of an isolated system is taken to be continuous,
$\widetilde{S}_{0}$ must be continuous. This ensures that the difference
$S_{0}(t)-\widetilde{S}_{0}$ is also continuous, which then proves that
$G(T_{0},P_{0},t)$ is a continuous function.

Since the system is an extremely small part of the isolated system, we have
\[
\left.  \left(  \frac{\partial S_{0}}{\partial\xi_{0}}\right)  \right\vert
_{\mathbf{Z}_{0}}=\left.  \left(  \frac{\partial\widetilde{S}}{\partial
\widetilde{\xi}}\right)  \right\vert _{\mathbf{Z}_{0}}=\frac{A_{0\xi}}{T_{0}%
}=0
\]
to a high degree of accuracy. Thus, we have
\[
\frac{d}{dt}\left[  S_{0}(E_{0},V_{0},\xi_{0},t)-\widetilde{S}(E_{0},V_{0}%
,\xi_{0})\right]  =\frac{\partial S_{0}}{\partial t}=\frac{dS_{0}}{dt}\geq0,
\]
where we have used $A_{0\xi}=0$, the constancy of $E_{0},V_{0}$ and the second
law. The above equation is valid even it $\xi_{0}$ is not constant, and
immediately proves Eq. (\ref{Gibbs_Free_Energy_Variation}).

Thus, we come to another very important conclusion, which we state as a theorem:

\begin{theorem}
\label{Theorem_Gibbs_free_energy_continuity}The Gibbs free energy
$G(T_{0},P_{0},t)$\ of the system is always a continuous function of time that
continuously decreases in time in accordance with Eq.
(\ref{Gibbs_Free_Energy_Variation}).
\end{theorem}

The above result does not require any knowledge of the state of the system
$\Sigma$. In particular, the continuity of $G(t)$ must remain valid even if
the processes going on within the system are chemical reactions, ergodicity
loss, chaos, turbulence, explosion, etc. or just normal quasi-static slow
processes going on in $\Sigma$. Thus, the continuity of $G(t)$ is valid when
the system is not in equilibrium with itself or with the medium. As time goes
on, $G(T_{0},P_{0},t)$ will continue to decrease towards the equilibrium value
$G(T_{0},P_{0})=E(T_{0},P_{0})-T_{0}S(T_{0},P_{0})+P_{0}V(T_{0},P_{0})$.

We now follow its consequences. Let us assume that the system is in
equilibrium with the medium at some temperature $T_{1}$ and pressure $P_{1}$
of the latter. The Gibbs free energy of $\Sigma$ in equilibrium is
\[
G(T_{1},P_{1})=E(T_{1},P_{1})-T_{1}S(T_{1},P_{1})+P_{1}V(T_{1},P_{1});
\]
obviously, equilibrium quantities do not have any explicit $t$-dependence. At
$t=0$, the system is brought in contact with a medium at the current
temperature $T_{0}$ and pressure $P_{0}$. As the energy, entropy and volume
have not had any time to change, the new Gibbs free energy at $t=0$ will be%
\[
G(T_{0},P_{0},t=0)=E(T_{1},P_{1})-T_{0}S(T_{1},P_{1})+P_{0}V(T_{1},P_{1}).
\]
This is true regardless of what changes may occur at $t>0$. It is clear that
the two Gibbs free energies are very different. Their difference is%
\begin{equation}
\Delta G(t=0)=-S(T_{1},P_{1})\Delta T+V(T_{1},P_{1})\Delta P\neq0,
\label{Differece_G(0)}%
\end{equation}
where $\Delta q=q_{\text{current}}-q_{\text{prior}}$ for any quantity $q$. The
difference $\Delta G(t=0)$ at $t=0$, however, has no physical significance for
the second law as the current isolated system with the medium at $T_{0},P_{0}$
is \emph{different} from the previous isolated system with the medium at
$T_{1},P_{1}$, while the second law refers to the behavior of a body in time
under fixed macroscopic conditions. We should also note that $\Delta
G(t=0)\rightarrow0$ as $\Delta T$ and $\Delta P$\ vanish simultaneously for
any system. Thus, if the macroscopic conditions do no change, the Gibbs free
energy undergoes no discontinuity at $t=0$.

For $t\geq0$, however, we are observing the same current isolated system, for
which Eq. (\ref{Gibbs_Free_Energy_Entropy_Relation}) holds so that the Gibbs
free energy $G$ of the system must vary \emph{continuously }$\ $for $t\geq0$
regardless of the kinds of processes going on inside $\Sigma_{0}$.

It is interesting to note that for infinitesimal changes in the temperature
and pressure, Eq. (\ref{Differece_G(0)}) reduces to%
\begin{equation}
dG(t=0)=-S(T_{1},P_{1})dT+V(T_{1},P_{1})dP,
\label{Gibbs_FreeEnergy_Differential}%
\end{equation}
a familiar expression from equilibrium thermodynamics.

If and when the internal equilibrium has been established in $\Sigma$, we can
introduce, as discussed in I, its \emph{instantaneous} fields and affinities,
such as the temperature $T\equiv T(t)$ and pressure $P\equiv P(t)$:%
\begin{equation}
\frac{1}{T(t)}=\frac{\partial S}{\partial E},\ \frac{P(t)}{T(t)}%
=\frac{\partial S}{\partial V},\frac{A_{\xi}(t)}{T(t)}=\frac{\partial
S}{\partial\xi}; \label{Eq_Conds}%
\end{equation}
see Eq. (\ref{Field_Affinity}). These are standard relations for the entropy
for a body under internal equilibrium and are used commonly in non-equilibrium
thermodynamics
\cite{Donder,deGroot,Prigogine,Guj-NE-I,Guj-NE-II,Gujrati-Symmetry}. To ensure
that the inequality in Eq. (\ref{Second_Law}) remains valid during relaxation,
i.e. during approach to equilibrium, we must have (see also \cite{Langer})%

\begin{equation}
T(t)\neq T_{0},\ P(t)\neq P_{0},A_{\xi}(t)\neq A_{0\xi}=0.
\label{Instantaneous_Parameters}%
\end{equation}
They become identical \emph{only} when equilibrium has been achieved. Thus, as
long as the relaxation is going on due to the absence of equilibrium, the two
inequalities must hold true; see I. Thus, we come to another important conclusions:

The use of the internal equilibrium is basically in the spirit of Onsager's
regression hypothesis \cite{Onsager}: Non-equilibrium relaxation is governed
by the same laws as the relaxation of spontaneous fluctuations occurring in an
equilibrium system.

We will assume that in an \emph{isobaric} cooling experiment,%
\begin{equation}
P(t)=P_{0}, \label{Mech_Equil}%
\end{equation}
which we will refer to as the existence of the \emph{mechanical equilibrium
}for the system. In this case, we find that
\begin{equation}
\frac{dS_{0}(t)}{dt}=\left(  \frac{1}{T}-\frac{1}{T_{0}}\right)  \frac{dH}%
{dt}\geq0. \label{Total_Entropy_Rate}%
\end{equation}
It can be shown on general grounds (see I) that in a cooling experiment under
mechanical equilibrium%
\begin{equation}
\frac{dH}{dt}<0 \label{Relaxation_Facts}%
\end{equation}
during relaxation in glasses. This is also seen experimentally.\ Thus, in such
a cooling experiment
\begin{equation}
T\geq T_{0}, \label{Temp_Relaxation}%
\end{equation}
the equality occurring only when equilibrium has been achieved. Furthermore,
in such cooling
\begin{equation}
\frac{dS}{dt}=\frac{1}{T}\frac{dH}{dt}. \label{Entropy_variation1}%
\end{equation}
during relaxation at fixed $T_{0}.$ The relaxation that occurs in the glass
originates from its tendency to come to thermal equilibrium during which its
temperature $T(t)$ varies with time; recall that we are considering a cooling
experiment under mechanical equilibrium. The relaxation process results in the
lowering of the corresponding Gibbs free energy, which is a consequence of the
second law in Eq. (\ref{Second_Law}) and is valid even when there is no
mechanical equilibrium. However, under mechanical equilibrium, the changes in
the enthalpy and entropy are in the same direction; see Eq.
(\ref{Entropy_variation1}). This then gives rise to the following

\begin{theorem}
\label{Theorem_Entropy_Relaxation}The lowering of $G(t)$ with time in an
isobaric experiment under mechanical equilibrium results in not only lowering
the enthalpy in a cooling experiment, as observed experimentally, but also the
entropy $S(t)$ during relaxation:%
\begin{equation}
dS/dt\leq0, \label{Entropy_variation_system}%
\end{equation}
as shown in Fig. \ref{Fig_EntropyLoss} for the conventional glass (CV) as it
approaches towards the supercooled liquid during relaxation.
\end{theorem}

\subsection{Gibbs fundamental relation}

It follows from Eqs. (\ref{Eq_Conds}) and (\ref{Eq_Conds_t}) that%
\begin{equation}
dS(t)=\frac{1}{T(t)}dE(t)+\frac{P(t)}{T(t)}dV(t)+\frac{A_{\xi}(t)}{T(t)}%
d\xi(t). \label{Gibbs_Fundamental_Relation}%
\end{equation}
This relation is known as the Gibbs fundamental relation. The first law of
thermodynamics is codified in the following differential:%
\begin{align}
dE(t)  &  =T(t)dS(t)-P(t)dV(t)-A_{\xi}(t)d\xi(t)\label{First_Law}\\
&  =T_{0}dS(t)-P_{0}dV(t)-A_{\text{S}}d\xi_{\text{S}}-A_{\text{V}}%
d\xi_{\text{V}}-A_{\xi}(t)d\xi(t), \label{First_Law_1}%
\end{align}
where the two new and different internal order parameters or degrees of
advancement
\[
\xi_{\text{S}}\equiv\frac{S(t)-S(\infty)}{S(0)-S(\infty)},\ \xi_{\text{V}%
}\equiv\frac{V(t)-V(\infty)}{V(0)-V(\infty)},
\]
are determined by the instantaneous entropy and volume governed by their
corresponding "affinities"%
\[
A_{\text{S}}\equiv-[T(t)-T_{0}]\left[  S(0)-S(\infty)\right]  ,\ A_{\text{V}%
}\equiv\lbrack P(t)-P_{0}]\left[  V(0)-V(\infty)\right]  ,
\]
thus, $\xi_{\text{S}}$,$\ \xi_{\text{V}}$ play the role similar to that of
internal variables $\xi,$ with $A_{\text{S}},A_{\text{V}}$\ playing the role
of their respective affinity. The differential of $G(t)$ and $H(t)$, see Eq.
(\ref{Free_Energies}), turns out to be%
\begin{equation}
dG(t)=-S(t)dT_{0}+V(t)dP_{0}-A_{\text{S}}d\xi_{\text{S}}-A_{\text{V}}%
d\xi_{\text{V}}-A_{\xi}(t)d\xi(t). \label{First_Law_Gibbs}%
\end{equation}
and
\begin{equation}
dH(t)=T_{0}dS(t)+V(t)dP_{0}-A_{\text{S}}d\xi_{\text{S}}-A_{\text{V}}%
d\xi_{\text{V}}-A_{\xi}(t)d\xi(t). \label{First_Law_Enthalpy}%
\end{equation}

For the isobaric case discussed above, we have $A_{\text{V}}=0,$ but
$A_{\text{S}}$ is normally non-zero. However, in almost all applications of
classical non-equilibrium thermodynamics to glasses to date that we are
familiar with, even $A_{\text{S}}$ is taken to be zero
\cite{Nemilov-Book,Gutzow-Book}:%
\begin{equation}
A_{\text{S}}=0. \label{Common_Approx}%
\end{equation}
This is equivalent to having%
\begin{equation}
T(t)=T_{0}, \label{Thermal_Equil}%
\end{equation}
which we will refer to here as the assumption of\emph{ thermal equilibrium}.
From Eq. (\ref{First_Law_Gibbs}), we see that $G$ cannot vary in time if we
have both thermal and mechanical equilibrium and if there are no internal
order parameters. Thus, there will be no relaxation unless we allow an order parameter:

\begin{theorem}
Under the assumption of thermal and mechanical equilibrium, internal order
parameters are required to describe relaxation observed in glasses.
\end{theorem}

Except in Sect. \ref{Marker_Stability}, we will not assume mechanical and
thermal equilibrium so as to remain as general in our discussion as possible.

\section{Non-equilibrium Entropy\label{Marker_NonEq-S}}

The results in the previous section are based on the thermodynamic concept of
the entropy for a body, not necessarily in equilibrium. Only the difference in
the entropy has any meaning as the entropy itself has no unique value, unless
supplemented by Nernst postulate \cite{Landau,Gujrati-Nernst}. This should be
contrasted with the statistical concept of the entropy originally due to Gibbs
\cite{Gibbs} and based on first principles, which provides a unique value of
the entropy for a given macrostate. Everyone believes that the Gibbs entropy
is the correct thermodynamic equilibrium entropy \cite{Landau}. The situation
with statistical interpretation of non-equilibrium entropy appearing in Eq.
(\ref{Second_Law}) may not be so clear, which we now analyze in this section.
The discussion in this section is very general and neither the medium nor the
system is assumed to be in internal equilibrium.

\subsection{Fundamental Axiom of Non-equilibrium Thermodynamics}

As thermodynamics is an experimental science, it requires \emph{several
measurements} on the body to obtain reliable results. To avoid any influence
of the possible changes in the body brought about by measurements, we instead
prepare a large number $\mathcal{N}$\ of samples or replicas under
\emph{identical macroscopic conditions }on which measurements are made. The
replicas are otherwise \emph{independent} of each other in that they evolve
independently in time. This is consistent with the requirement that different
measurements should not influence each other. The samples are prepared so that
the probability of a sample in microstate $j$ is $p_{j}(\mathbf{Z}(t),t)$,
which we simply write as $p_{j}(t)$; it is a \emph{continuous} function of the
state variable $\mathbf{Z}(t)$ and of $t$ for a macroscopically large body. We
now state the fundamental axiom of thermodynamics as proposed in
\cite{Gujrati-Symmetry}:

\begin{quote}
\textsc{Fundamental Axiom }\emph{The thermodynamic behavior of a system is not
the behavior of a single sample, but the average behavior of a large number of
independent samples, prepared identically under the same macroscopic
conditions at time}\textsl{\ }$t=0$\textsl{. }
\end{quote}

Such an approach is standard in equilibrium statistical mechanics
\cite{Landau,Tolman,Huang,Becker}, but it must also apply to systems not in
equilibrium as it is required for the reliability of measurements. For
non-equilibrium systems, this averaging must be carried out by ensuring that
all samples have been prepared with identical history. This is obviously not
an issue for systems in equilibrium. We refer the reader to a great discussion
about the status of statistical mechanics and its statistical nature in Sect.
25 by Tolman \cite{Tolman}; see also the last paragraph on p. 106 in Jaynes
\cite{Jaynes}. This point has been recently reviewed in
\cite{GujratiResidualentropy,Gujrati-Symmetry}.

The average over these samples of some thermodynamic quantity $X$ then
determines the thermodynamic average quantity $\overline{X}$ for the body
\begin{equation}
\overline{X}(t)\equiv\sum_{j=1}^{W}p_{j}(t)X_{j}, \label{Av_X}%
\end{equation}
where $X_{j}$ is the value of $X$ in the $j$th microstate of the body, and $W$
is the number of its distinct microstates.

\subsection{General Formulation of Entropy: Isolated system}

\subsubsection{Gibbs Formulation of Non-equilibrium entropy}

We begin by considering an isolated system $\Sigma_{0}$, which need not be in
equilibrium. Gibbs \cite{Gibbs}, Tolman \cite{Tolman} and Rice and Gray
\cite{Rice} among others discuss at length the non-equilibrium entropy using
the Gibbs formulation; see also \cite{GujratiResidualentropy,Gujrati-Symmetry}%
. It is given by the negative average of%
\[
\eta\equiv\ln p,
\]
what Gibbs \cite{Gibbs} calls the \emph{index of probability:}
\begin{equation}
S_{0}(t)\equiv-\overline{\ln p}\equiv-%
%TCIMACRO{\tsum \limits_{\alpha}}%
%BeginExpansion
{\textstyle\sum\limits_{\alpha}}
%EndExpansion
p_{\alpha}(t)\ln p_{\alpha}(t),\ \ \ \ \
%TCIMACRO{\tsum \limits_{\alpha}}%
%BeginExpansion
{\textstyle\sum\limits_{\alpha}}
%EndExpansion
p_{\alpha}(t)\equiv1, \label{Conventional_Entropy}%
\end{equation}
where $p_{\alpha}(t)$ is the probability of the $\alpha$th microstate at time
$t$. It is shown below that it follows from Eq. (\ref{Av_X}). As the
microstate probability $p_{\alpha}(t)$ is a continuous function of its
arguments $\mathbf{Z}_{0}$ and $t$, the entropy $S_{0}(t)\equiv S_{0}%
(\mathbf{Z}_{0},t)$ is also a continuous function of its arguments
$\mathbf{Z}_{0}$ and $t$. It is straightforward to establish
\cite{Tolman,Rice} that this entropy satisfies Eq. (\ref{Second_Law}). The
identification of the entropy with the negative of the Boltzmann $H$-function
\cite[see p. 561]{Tolman}, the latter describing a non-equilibrium state,
should leave no doubt in anyone's mind that the Gibbs formulation of the
entropy can be applied equally well to an equilibrium or a non-equilibrium
system in isolation.

For a generic body, we will write the continuous entropy as%
\begin{equation}
S(t)\equiv-\overline{\ln p}\equiv-\sum_{j}p_{j}(t)\ln p_{j}(t),\ \ \ \
%TCIMACRO{\tsum \limits_{j}}%
%BeginExpansion
{\textstyle\sum\limits_{j}}
%EndExpansion
p_{j}(t)\equiv1, \label{Conventional_Entropy_System}%
\end{equation}
where $p_{j}(t)$ is the probability of the $j$th microstate of the body at
time $t$. The justification for the entropy expression in Eq.
(\ref{Conventional_Entropy_System}) for an open system, its continuity and its
connection with the second law will be given below.

There are two different ways to understand the above entropy formulation.

\subsubsection{Ensemble Interpretation-Dynamics being Irrelevant}

\ We now prove that the entropy is a statistical average given in Eq.
(\ref{Av_X}). We consider a large number $\mathcal{N=C}W_{0}(\mathbf{X}_{0})$
of independent \emph{replicas} or \emph{samples} of the isolated system, with
$\mathcal{C}$\ some large constant integer. Let there be $\mathcal{N}_{\alpha
}$ samples in the microstate $\alpha.$ The probability of a sample in
microstate $\mathcal{\alpha}$ is then%
\begin{equation}
p_{\alpha}=\mathcal{N}_{\alpha}/\mathcal{N}. \label{sample_probability}%
\end{equation}
The ensemble average of $\mathbf{Z}_{0}$\ over these samples is given by%
\begin{equation}
\overline{\mathbf{Z}}_{0}=\frac{1}{\mathcal{N}}%
%TCIMACRO{\tsum \limits_{\alpha=1}^{W_{0}(\mathbf{X}_{0})}}%
%BeginExpansion
{\textstyle\sum\limits_{\alpha=1}^{W_{0}(\mathbf{X}_{0})}}
%EndExpansion
\mathcal{N}_{\alpha}\mathbf{Z}_{0\alpha}=%
%TCIMACRO{\tsum \limits_{\alpha=1}^{W_{0}(\mathbf{X}_{0})}}%
%BeginExpansion
{\textstyle\sum\limits_{\alpha=1}^{W_{0}(\mathbf{X}_{0})}}
%EndExpansion
p_{\alpha}\mathbf{Z}_{0\alpha}, \label{Ensemble_Average}%
\end{equation}
where $\mathbf{Z}_{0\alpha}$ is the value of $\mathbf{Z}_{0}$ in the
microstate $\alpha$. This is identical to the statistical average given in Eq.
(\ref{Av_X}).

The number of ways $\mathcal{W}$ to arrange the $\mathcal{N}$ samples into
$W_{0}$ distinct microstates so that there are $\mathcal{N}_{\alpha}$ samples
in microstate $\alpha$ is given by%
\[
\mathcal{W\equiv}\frac{\mathcal{N}!}{%
%TCIMACRO{\tprod \limits_{\alpha}}%
%BeginExpansion
{\textstyle\prod\limits_{\alpha}}
%EndExpansion
\mathcal{N}_{\alpha}!}.
\]
Taking its natural log to obtain an additive quantity
\begin{equation}
\mathcal{S}\equiv\ln\mathcal{W}, \label{Ensemble_entropy_Formulation}%
\end{equation}
and using Stirling's approximation for the factorials, we see easily that
$\mathcal{S}$ per sample can be written as
\[
\frac{\mathcal{S}}{\mathcal{N}}=-%
%TCIMACRO{\tsum \limits_{\alpha=1}^{W_{0}(\mathbf{X}_{0})}}%
%BeginExpansion
{\textstyle\sum\limits_{\alpha=1}^{W_{0}(\mathbf{X}_{0})}}
%EndExpansion
p_{\alpha}\ln p_{\alpha},
\]
where $p_{\alpha}$ is given in Eq. (\ref{sample_probability}). Thus,
$\mathcal{S}/\mathcal{N}$ is the ensemble average of $-\ln p_{\alpha}$; thus,
it is nothing but the entropy $S_{0}$ of the body given in Eq.
(\ref{Conventional_Entropy}). Again, the continuity of $\mathcal{S}%
/\mathcal{N}$ follows directly from the continuity of $p_{\alpha}$.

The maximum possible value of the entropy for fixed $\mathbf{X}_{0}$ is
\begin{equation}
S_{0,\text{max}}(\mathbf{X}_{0})=\ln W_{0}(\mathbf{X}_{0}),
\label{S_Boltzmann}%
\end{equation}
which occurs if and only if all microstates are \emph{equally probable}:%
\[
p_{\alpha}(t)\rightarrow1/W_{0}(\mathbf{X}_{0}),\text{~~\ \ }\forall\alpha
\in\Gamma_{0}.
\]
This maximum value is the Boltzmann entropy of the system. This certainly
occurs as $t\rightarrow\infty$ for most systems. Then the system is said to be
in equilibrium. It is evident that the Gibbs formulation in Eq.
(\ref{Conventional_Entropy}) is more general than the Boltzmann formulation in
Eq. (\ref{S_Boltzmann}), as the former contains the latter as a special limit.
In equilibrium, $\mathcal{N}_{\alpha}=\mathcal{C}$ for all $\alpha$, so that
$p_{\alpha}=1/W_{0}$.

\subsubsection{Temporal Interpretation-Dynamics being Relevant}

Another way to interpret Eq. (\ref{Conventional_Entropy}) is as follows, which
is quite standard, at least in developing the kinetic theory of gases. We
consider the time-evolution of an isolated system
\cite{GujratiResidualentropy,Gujrati-Symmetry}, starting at $t=0$ from some
initial microstates $\alpha_{0}$. Then $p_{\alpha}(t)$ at some later time $t$
represents the frequency with which the $\alpha$th microstate has occurred
\emph{during} this time interval. This is the temporal definition of the
probability \cite{Landau}, the traditional way of introducing this
probability. All these microstates belong to the slice $\Gamma_{0}$ of the
microstate space,\ consisting of fixed $\mathbf{X}_{0},$ but different
$\mathbf{\xi}_{0}(t)$. At $t=0$, $p_{\alpha}(0)=\delta_{\alpha,\alpha
_{0}\text{ }}$, where $\delta$ is the Kronecker delta, and $S_{0}(0)=0$. With
time, this entropy will continue to increase until it reaches its maximum for
fixed $\mathbf{X}_{0}$. Let $W_{0}(\mathbf{X}_{0})$ denote the number of
distinct microstates in the slice $\Gamma_{0}$.

For whatever reasons, if it happens that the system is confined to, or has
visited during certain time interval $\tau$ (such as the observation time
$\tau_{\text{obs}}$), only a part $\Gamma_{0}^{\prime}\subset\Gamma_{0}$ of
the above microstate space slice, then%
\begin{equation}
\ p_{\alpha}(\tau)=0,\ \ \ \ \forall\alpha\notin\Gamma_{0}^{\prime},
\label{OutsideMicrostates}%
\end{equation}
even though these microstates corresponds to the same $\mathbf{X}_{0}$. This
happens because these microstates have not been visited yet. Let
$W_{0}^{\prime}(\mathbf{X}_{0},\tau)<W_{0}(\mathbf{X}_{0})$ denote the number
of distinct microstates in $\Gamma_{0}^{\prime}$. Then, the conventional
entropy in Eq. (\ref{Conventional_Entropy}) with the above condition in Eq.
(\ref{OutsideMicrostates}) would be strictly bounded
\[
S_{0}(\mathbf{X}_{0},\tau)\equiv-%
%TCIMACRO{\tsum \limits_{\alpha\in\Gamma_{0}^{\prime}}}%
%BeginExpansion
{\textstyle\sum\limits_{\alpha\in\Gamma_{0}^{\prime}}}
%EndExpansion
p_{\alpha}(t)\ln p_{\alpha}(t)\leq\ln W_{0}^{\prime}(\mathbf{X}_{0},\tau);
\]
the equality occurs if and only if all these microstates in $\Gamma
_{0}^{\prime}$\ happen to be \emph{equally probable}:
\begin{equation}
p_{\alpha}(\tau)=1/W_{0}^{\prime}(\mathbf{X}_{0},\tau),\text{~~\ \ }%
\forall\alpha\in\Gamma_{0}^{\prime}.\ \ \ \ \label{Eq_Limiting_Value_p}%
\end{equation}
Again we notice the generality of the Gibbs formulation in Eq.
(\ref{Conventional_Entropy}) over the Boltzmann formulation in Eq.
(\ref{S_Boltzmann}).

The ensemble approach is also very important from an experimental point of
view. At high temperatures, where dynamics is very fast, it is well known that
it agrees with the temporal formulation. However, at low temperatures, where
dynamics becomes sluggish as in a glass, the temporal entropy can be
misleading. However, the ensemble entropy still gives the correct value. This
point has been discussed in \cite{Gujrati-Symmetry}.

\subsection{General Formulation of Entropy: Open
system\label{Marker_Entropy_Open_System}}

Let us now consider our system $\Sigma$ with at least one extensive observable
that is fixed \cite{Gujrati-Fluctuations}, which we take to be the number of
particles $N$. As usual, the system is a small part of the isolated system
$\Sigma_{0}$. We wish to determine the entropy of $\Sigma$ in terms of its
microstates, which are indexed by $i$. Theses microstates correspond to all
possible allowed energies and volumes. We use $\widetilde{\alpha}$ to denote
the microstates of $\widetilde{\Sigma}$ with a fixed number of particles
$\widetilde{N}$. A specification of the microstates $i$ and $\widetilde
{\alpha}$ gives a unique microstate specification $\alpha$. Hence, the number
of microstates $W_{0}\ $of the $\Sigma_{0}$\ is the product $W\widetilde{W}$,
where $W$ and $\widetilde{W}$ are respectively the number of microstates of
$\Sigma$\ and $\widetilde{\Sigma}$.\ Because of the smallness of $\Sigma$
relative to $\Sigma_{0}$, which results in the \emph{quasi-independence} of
the system and the medium to a very high degree of accuracy, we have in terms
of their probabilities
\[
p_{\alpha}(t)=p_{i}(t)p_{\widetilde{\alpha}}(t).
\]
As usual, these probabilities are continuous functions of their arguments.
Now, using $\ln p_{i}(t)p_{\widetilde{\alpha}}(t)=\ln p_{i}(t)+\ln
p_{\widetilde{\alpha}}(t)$, and the sum rule%
\[%
%TCIMACRO{\tsum \limits_{\widetilde{\alpha}}}%
%BeginExpansion
{\textstyle\sum\limits_{\widetilde{\alpha}}}
%EndExpansion
p_{\widetilde{\alpha}}(t)=1,\ \
%TCIMACRO{\tsum \limits_{i}}%
%BeginExpansion
{\textstyle\sum\limits_{i}}
%EndExpansion
p_{i}(t)=1,
\]
we find that
\[
S_{0}(t)\equiv-%
%TCIMACRO{\tsum \limits_{i}}%
%BeginExpansion
{\textstyle\sum\limits_{i}}
%EndExpansion
p_{i}(t)\ln p_{i}(t)-%
%TCIMACRO{\tsum \limits_{\widetilde{\alpha}}}%
%BeginExpansion
{\textstyle\sum\limits_{\widetilde{\alpha}}}
%EndExpansion
p_{\widetilde{\alpha}}(t)\ln p_{\widetilde{\alpha}}(t),
\]
where the two terms in the above equations represent the entropies of the
system and the medium%
\begin{equation}
S(t)=-%
%TCIMACRO{\tsum \limits_{i}}%
%BeginExpansion
{\textstyle\sum\limits_{i}}
%EndExpansion
p_{i}(t)\ln p_{i}(t),\ \ \ \widetilde{S}(t)=-%
%TCIMACRO{\tsum \limits_{\widetilde{\alpha}}}%
%BeginExpansion
{\textstyle\sum\limits_{\widetilde{\alpha}}}
%EndExpansion
p_{\widetilde{\alpha}}(t)\ln p_{\widetilde{\alpha}}(t). \label{Entropies}%
\end{equation}
Note that our entire discussion never puts any restriction on the kind of
actual processes going on during approach to equilibrium. Hence, whether there
is any glass transition, any loss of ergodicity loss, etc., has no bearing on
the formulation of entropies in Eq. (\ref{Entropies}) or our other general
results presented in the previous section.

Let us imagine isolating the medium by disconnecting it from the system. Its
entropy is given by $\widetilde{S}(t)\equiv\ \widetilde{S}(\widetilde
{\mathbf{Z}}(t),t)$. The isolated medium is similar to the isolated system
$\Sigma_{0}$, whose entropy was assumed to be continuous; see the discussion
immediately after Eq. (\ref{Second_Law}). Thus, $\ \widetilde{S}(t)$ must also
be a continuous function. The continuity of $S_{0}(t)$ and of $\ \widetilde
{S}(t)$ then immediately leads to the continuity of the entropy $S(t)$ of the
system. This result is also consistent with the continuity of microstate
probabilities $p_{i}(t)$. The continuity of $S(t)$ is independent of the kinds
of processes that are going on within it. This discussion then leads us to the
following important theorem:

\begin{theorem}
\label{Theorem_Continuous_System_Entropy}The entropy $S(\mathbf{Z}(t),t)$ of
an open system is a continuous function of its arguments under all conditions.
In particular, it is also continuous when the medium is in internal equilibrium.
\end{theorem}

The above derivation not only justifies Eq. (\ref{Entropies_Sum}) or Eq.
(\ref{Total_Entropy}) for an isolated system, but also identifies the proper
non-equilibrium entropy for our open system $\Sigma$, which is always a
continuous function of its extensive arguments and $t.$ The expression for $S$
in terms of microstate probabilities in Eq. (\ref{Entropies}) is identical in
form to what one would use for an equilibrium system \cite{Landau}, except
that the microstate probabilities are now explicit functions of time $t$. This
entropy then determines the Gibbs free energy and enthalpy in Eq.
(\ref{Free_Energies}) of the system. The second law in Eq. (\ref{Second_Law})
for the isolated system $\Sigma_{0}$ is now expressed in terms of the Gibbs
free energy for the system $\Sigma$ and is given in Eq.
(\ref{Gibbs_Free_Energy_Variation}), provided that the medium is taken to be
in internal equilibrium. The fact that $\Sigma$ is in a medium $\widetilde
{\Sigma}$ is reflected in the fact that the $G(t)$ and $H(t)$ depend on
$T_{0},P_{0}$ of the medium, and not on $T(t),P(t)$ and $A_{\xi}(t)$\ of the
system. Recall that the system may have no well-defined temperature, pressure,
etc. unless the system is in internal equilibrium. This is important to
emphasize as the validity of the second law cannot depend on the establishment
of internal equilibrium in the system. Thus, the entropy formulation in Eq.
(\ref{Entropies}) is very general.

Notice that the ensemble interpretation of the entropy in a body is due to the
choice of samples with the probability $p_{i}$; the \emph{dynamics} within the
body never appears in the discussion. Thus, such an interpretation is quite
useful when the dynamics in a system becomes very sluggish, such as in a glass.

\section{Principles of Non-equilibrium
Thermodynamics\label{Sect_Thermodynamic_Principles}}

Based on the discussion above and the \textsc{Fundamental Axiom}, we can
formulate four fundamental principles of non-equilibrium thermodynamics
\cite{GujratiResidualentropy}. These principles are well known and accepted in
equilibrium thermodynamics. We now discuss why they should also be applicable
to \emph{slowly} evolving metastable states such as glasses in which we are
interested in this work. We will assume that the system of interest is in
internal equilibrium, but not in equilibrium with the medium.

\begin{enumerate}
\item The first one is the \emph{principle of additivity}, according to which
the total entropy or other extensive quantities can be obtained by a sum over
different macroscopic parts of the body. Each part must be large enough so
that the usual argument that their surface effects can be neglected as
thermodynamically unimportant is valid so that the parts become
\emph{quasi-independent} \cite{Landau,Guj-NE-II}.
\end{enumerate}

This principle is consistent with what one must do for any measurement, which
requires verification by performing the measurement many times over on
different samples and performing an average over these samples. Each
measurement must be performed under identical macroscopic conditions. We now
argue that different quasi-independent parts of the body represent different
samples of the body prepared under identical conditions. Consider some body
$\Sigma^{\prime}$ with $N^{\prime}$\ particles, and imagine a much larger body
$\Sigma$ that can be divided into a large number of parts $\Sigma_{k}^{\prime
},k=1,2,\cdots$, each part with $N^{\prime}$ particles so that they are the
same size as the body $\Sigma^{\prime}$. Let us imagine $\Sigma$ itself to be
a very small part of an isolated system $\Sigma_{0},$ so that various
$\Sigma_{k}^{\prime}$ and $\Sigma$ do not affect the internal equilibrium of
the much larger medium $\widetilde{\Sigma}$. In particular, $T_{0},P_{0}$ and
$A_{0\xi}=0$\ of $\widetilde{\Sigma}$ are unaffected by $\Sigma_{k}^{\prime}$
and $\Sigma$. We can imagine each $\Sigma_{k}^{\prime}$ to be in a medium
$\widetilde{\Sigma}_{k}$, where the latter is composed of the medium
$\widetilde{\Sigma}$ and the remainder of the system $\Sigma$ obtained by
taking out the part $\Sigma_{k}^{\prime}$ under investigation. Because of the
smallness of $\Sigma$ relative to $\widetilde{\Sigma}$, the new medium also
has the same $T_{0},P_{0}$ and $A_{0\xi}=0$\ as the medium $\widetilde{\Sigma
}$. Thus, each part $\Sigma_{k}^{\prime}$ experiences the same $T_{0},P_{0}$
and $A_{0\xi}=0$ as any other part: all parts experience the same macroscopic
conditions. Because of this, different parts $\Sigma_{k}^{\prime}$ are nothing
but the preparation of the body $\Sigma^{\prime}$ many times over under
identical conditions specified by the same $T_{0},P_{0}$ and $A_{0\xi}=0$.
Thus, averaging any quantity over different samples is equivalent to averaging
over different parts $\Sigma_{k}^{\prime}$ of $\Sigma$.\ This principle of
additivity must apply to any body even if it is not in equilibrium.
Consequently, the value of any extensive quantity over the entire body must be
a sum over its various parts, regardless of whether the body is in equilibrium
or not.

\begin{enumerate}
\item[2.] The second principle is the \emph{principle of reproducibility}%
,\emph{ }according to which the ensemble average is equal to the average of
the experimental values, also called the thermodynamic average.
\end{enumerate}

This principle follows from the \textsc{Fundamental Axiom }according to which
thermodynamics, whether equilibrium or time-dependent, requires several
measurements on the system to obtain reliable results. To avoid any influence
of the possible changes in the system brought about by measurements, we can
instead prepare a large number of samples under identical macroscopic
conditions. This is consistent with the requirement that different
measurements should not influence each other. Then the principle follows
immediately from Eq. (\ref{Ensemble_Average}).

\begin{enumerate}
\item[3.] The third principle is the \emph{principle of continuity and
uniqueness }, which states that the Gibbs free energy is continuous, and hence
a unique (single-valued) function of its arguments.
\end{enumerate}

For equilibrium states, this principle is certainly valid. We extend it
non-equilibrium systems. The principle follows from Conclusion 2. As said
there, its validity is independent of the kinds of processes going on within
the system $\Sigma$. In the absence of uniqueness, different experimentalists
will have no way to effectively communicate their results and no scientific
investigation can be carried out. This is a very important requirement if the
second law for an open system in the form of Eq.
(\ref{Gibbs_Free_Energy_Variation}) has to make any sense. The existence of
the Gibbs free energy, as demonstrated in Sect. \ref{Marker_Gibbs_Free_Energy}
does not require the system be at least in internal equilibrium; it only
requires the medium to be in internal equilibrium. Therefore, it is most
certainly also valid when the system is in internal equilibrium.

\begin{enumerate}
\item[4.] The last principle is the \emph{principle of stability}, according
to which the heat capacity, compressibility, etc. must remain non-negative for
the system to remain stable.
\end{enumerate}

We have already discussed the non-negativity of the heat capacity in Eqs.
(\ref{Heat_Capacity_Stability}-\ref{Heat_Capacity_Stability_1}). Similar
conditions also apply to compressibility and other response functions.

\section{Entropy as a Macrostate Property and Component
Confinement\label{Sect_Confinement}}

Can there anything be wrong with the original argument that suggests entropy
reduction due to confinement as noted in Sect. \ref{Marker_Introduction}?
After all, if the system is confined to just one component, should not the
confinement entropy be zero? What can be wrong with such a simple argument? To
answer this question, we need to understand the concept of the statistical
entropy, to which we now turn. More details can be found in a recent review
\cite{Gujrati-Symmetry}.

\subsection{Non-interacting Ising Spins: Ensemble and Temporal Descriptions}

Let us consider an ensemble of $N$ \emph{non-interacting} Ising spins located
at the sites of a lattice with $N$ sites. All possible $W=2^{N}$ microstates
have the same energy $E=0$, and each microstate $\mathcal{I}_{j}$ has the same
\textit{a priori} probability $p_{j}=1/W$. The spin macrostate $\mathbf{I}$,
on the other hand,\ is only specified by the number of up and down spins, and
not the sequence of spin states on the lattice. One can relate this problem to
picking $N$ balls with replacement from an urn containing equal number of
balls of two colors. After picking a ball, we must place it back in the urn.
The sequence of the colors of the $N$ balls in \emph{time} with replacement
determines a microstate $\mathcal{B}_{j}$ of the $N$ balls. A ball macrostate
$\mathbf{B}$ is determined by specifying only the number of balls of each
color, but not the actual sequence of colors. We can identify the state of the
$k$th spin in $\mathcal{I}_{j}$ with the color of the ball picked at the $k$th
attempt in $\mathcal{B}_{j}$. Then the two problems are identical, except that
we are considering an ensemble of the Ising spins, whereas we are considering
a temporal description for the balls. Indeed, in the latter case, we have a
temporal evolution of the state of a single ball in time. Now, there is no
dynamics that allows a ball microstate $\mathcal{B}_{j}$ to change into
another ball microstate $\mathcal{B}_{j}^{\prime}$. There may be some dynamics
that could change a spin microstate $\mathcal{I}_{j}$ to another microstate
$\mathcal{I}_{j}^{\prime}$. Whether there is any microscopic dynamics
specified or not is irrelevant in determining the entropy for the spins.

The entropy for both systems, spins and balls, is
\[
S=N\ln2.
\]
It is also the maximum entropy, indicating that we are dealing with an
equilibrium state. This entropy will be the same even if there were no
dynamics changing the spins, such as at absolute zero in classical mechanics.
When a system is confined to one of disjoint components of the phase space
from which it cannot escape, we also encounter a situation with the absence of
a dynamics, the dynamics which takes the system from one component to another
\cite{Palmer,GujratiResidualentropy,Gujrati-Symmetry}. Accordingly, every
realization of the balls or spins will remain in its microstate forever, just
like a glassy system which remains confined to a disjoint component
\cite{Nemilov-Book,Gutzow-Book,Palmer,GujratiResidualentropy,Gujrati-Symmetry}
due to kinetic freezing. However, as the \textsc{Fundamental Axiom} states,
the entropy is an \emph{average} quantity obtained by an average over
\emph{all} samples or microstates, as shown in Eq.
(\ref{Conventional_Entropy_System}). This is equivalent to saying that
\emph{the entropy is determined by the macrostate}, which represents a
collection of microstates, each with certain \emph{a priori} probability. The
entropy has a contribution from all of these microstates. It is not the
property of a single microstate. This point should be very clear form the
derivation of relating $\mathcal{S}/\mathcal{N}$ in Eq.
(\ref{Ensemble_entropy_Formulation}) with the entropy. The evaluation of
$\mathcal{S}$ is due to the choice of various samples; the dynamics is not
part of the derivation. As a consequence, the entropy is unaffected by whether
microstates have any dynamics for change or not in time. One only needs to
know the probability distribution for the microstates.

For a kinetically frozen glass, different samples will correspond to a glass
confined to any one of the components
\cite{Palmer,GujratiResidualentropy,Gujrati-Symmetry}. We have no information
as to which component a sample is frozen into. Thus, the entropy must be
obtained by averaging over these different samples or disjoint components. We
cannot just consider one particular sample, since the entropy is not the
property of a single component.

\subsection{Probability Collapse and Entropy
Reduction\label{marker_Prob_Collapse}}

Is it possible to justify that the entropy is determined by considering just
one of the components in which the system is confined? If such a determination
is possible, it must be because the entropy discontinuously decreases when the
component confinement occurs.

The reduction in entropy is most certainly possible if we perform a
measurement to identify the microstate in which a given sample is. To identify
a particular microstate requires complete information about the body, which is
ordinarily unfeasible. In order to identify any particular microstate, we need
to perform a very special kind of measurement, which we will call a
\emph{microstate measurement} \cite{Gujrati-Symmetry}, that provides us with
the complete information about the body in its current microstate $j_{0}$. For
the Ising model, this requires \emph{determining} the orientations of each of
the $N$ spins. After the microstate measurement, we know with certainty which
microstate the system is in. Before the measurement, a given sample is known
to be in one of the microstates. The probability $p_{0}$ that the sample is in
microstate $j_{0}$ is $p_{0}=1/W.$\ This probability changes
\emph{discontinuously} from $p_{0}=1/W$ to $p_{0}=1$ immediately after the
measurement. The effect of the microstate measurement is to also reduce the
probabilities of all other microstates $j^{\prime}\neq j_{0}$ to
$p_{j^{\prime}}=0$ for this sample. Thus,
\[
p_{j}=\delta_{j,j_{0}},
\]
immediately after the measurement. We will speak of \emph{probability collapse
}to indicate this change in the probability brought about by the microstate
measurement in this work. The entropy also vanishes in an abrupt fashion
immediately after the measurement from the initial value of $\ln W$ in
accordance with complete certainty about the body.

While quite an appealing\ argument for the justification, it overlooks two
important facts:

\begin{enumerate}
\item In experimental glass transition, no such measurement is ever made that
identifies precisely which component the glass is frozen in. Such a
measurement will tell us precisely the positions and momenta of all the $N$
particles which then allows us to decipher which particular component the
glass is in.

\item Because of the lack of such a measurement, we must determine the entropy
by averaging over all components as shown in Eq. (\ref{S_components}) by
considering all possible samples with the probability distribution
$p_{\lambda}$ for the components.
\end{enumerate}

\subsection{System Confined in a Component:\ The Residual
Entropy\cite{Gujrati-Symmetry}}

In real glass formation, no probability collapse occurs as no microstate
measurement is ever performed to identify the particular component the sample
is in. The sample still has the same probability to be in any of the
$\mathcal{C}$ components it had before the confinement occurs. The confinement
does not alter the probability distribution. All we know for sure is that a
glass sample is in any one of the $\mathcal{C}$ components. We do not know the
actual component it is in. This situation is identical to the die which is
known to be in one of the six possibilities when it is concealed by the cup,
or to the non-interacting Ising spins discussed earlier. As long as the die is
concealed, the probability of an outcome of an unloaded die remains $1/6$,
regardless of what is on the top face. Even if the outcome cannot change when
we lift the cup, the outcome is most certainly not certain when concealed by
the cup; see \cite{Gujrati-Symmetry} for more details on this point. The
information that the glass has been formed is most certainly not equivalent to
knowing precisely the particular component in which the glass is trapped. The
latter will require the collapse of the original probability distribution by a
microstate measurement. The entropy is obtained by taking the average over all
the samples, and not only one glass sample, in accordance with the
\textsc{Fundamental Axiom}.

Let us now follow this line of reasoning. The entropy $S(t)$ can be written as
a sum over all the components, indexed by $\lambda=1,2.\cdots,\mathcal{C}$%
\[
S(t)=\sum\limits_{\lambda=1}^{\mathcal{C}}\sum\limits_{j_{\lambda}%
}(-p_{j_{\alpha}}\ln p_{j_{\alpha}}),\ \ \sum\limits_{\alpha=1}^{\mathcal{C}%
}\sum\limits_{j_{\alpha}}p_{j_{\alpha}}=1,
\]
where $j_{\lambda\text{ }}$represents one of the microstates in the component
$\lambda$, the first sum is over all the components and the second sum is over
all microstates in each component. The above way of writing is an identity and
applies under all cases. Thus, it applies whether confinement occurs or not.
Introducing%
\[
p_{\lambda}\equiv\sum\limits_{j_{\lambda}}p_{j_{\lambda}},\sum\limits_{\lambda
=1}^{\mathcal{C}}p_{\lambda}=1,
\]
we can rewrite $S(t)~$as follows:%
\begin{equation}
S(t)=\sum\limits_{\lambda=1}^{\mathcal{C}}p_{\lambda}S_{\lambda}%
(t)+S_{\mathcal{C}}(t), \label{S_components}%
\end{equation}
where
\[
S_{\lambda}(t)\equiv\sum\limits_{j_{\lambda}}(-\frac{p_{j_{\lambda}}%
}{p_{\lambda}}\ln\frac{p_{j_{\lambda}}}{p_{\lambda}})
\]
denotes the \emph{component} entropy in a given component $\lambda$, and%
\[
\ S_{\mathcal{C}}(t)\equiv\sum\limits_{\lambda=1}^{\mathcal{C}}(-p_{\lambda
}\ln p_{\lambda})
\]
denotes the \emph{confinement }entropy due to various components. For an
unbiased sampling of the components, $\ S_{\mathcal{C}}(t)$ must be at its
maximum, which requires
\begin{equation}
p_{\lambda}=1/\mathcal{C}. \label{p_component_unbiased}%
\end{equation}
For the case of the die, $S_{\mathcal{C}}=\ln6$ and $S_{\lambda}(t)=0$ to give
$S=\ln6,$ as noted earlier.

In real glasses, the entropy $S_{\lambda}(t)$ vanishes as the glass approaches
absolute zero. In this case, the entropy at absolute zero reduces to
$S_{\mathcal{C}}(t)$, which is what is customarily called the residual entropy
$S_{\text{R}}$ \cite{Nemilov-Book,Gutzow-Book}. We thus define the residual
entropy as
\begin{equation}
S_{\text{R}}\equiv-\sum\limits_{\lambda=1}^{\mathcal{C}}p_{\lambda}\ln
p_{\lambda}; \label{Residual_Entropy}%
\end{equation}
when the system explores various components without any bias so that Eq.
(\ref{p_component_unbiased}) holds, the residual entropy reduces to%
\begin{equation}
S_{\text{R}}=\ln\mathcal{C}. \label{Residual_Entropy_0}%
\end{equation}

\subsection{Residual Entropy of Subsystems and Calorimetric
Measurements\label{Sec.SubsystemEntropy}}

Another way to understand Eq. (\ref{Residual_Entropy}) is to recall the
additive of the entropy. We imagine dividing the system $\Sigma$ into several
macroscopically large but equal parts of size $N^{^{^{\prime}}},$ these parts
representing many samples of a smaller system $\Sigma^{^{\prime}}$. Each part
now represents a glass trapped in a component $\lambda^{\prime}$ of the
smaller system $\Sigma^{^{\prime}}$. Let $\mathcal{C}^{\prime}$ denote the
number of disjoint components for $\Sigma^{\prime}$. Then the number of
disjoint components $\mathcal{C}$ for $\Sigma$ is given by%
\[
\mathcal{C}=(\mathcal{C}^{\prime})^{N/N^{^{^{\prime}}}},
\]
where $N/N^{^{\prime}}$\ denotes the number of $\Sigma^{\prime}$ parts in
$\Sigma$. Again, we cannot be sure of which component $\lambda^{\prime}$ each
part is trapped in. The components appear with probabilities $p_{\lambda
^{\prime}}$ for any part so that%
\[
\ S_{\text{R}}^{\prime}\equiv\sum\limits_{\lambda^{\prime}=1}^{\mathcal{C}%
^{\prime}}(-p_{\lambda^{\prime}}\ln p_{\lambda^{\prime}})
\]
for each part. The probability of $\Sigma$ is obtained by adding this entropy
over all parts so that%
\[
S_{\text{R}}\equiv\frac{N}{N^{\prime}}S_{\text{R}}^{\prime}.
\]
If the components appear with no bias, then
\begin{equation}
S_{\text{R}}=\frac{N}{N^{\prime}}\ln\mathcal{C}^{\prime}, \label{Res_S_Parts}%
\end{equation}
which coincides with the result in Eq. (\ref{Residual_Entropy_0}) for $\Sigma
$. The additivity principle clearly shows that the residual entropy for
$\Sigma$ is not zero.

It is evident now that to conclude that the residual entropy has vanished just
because a sample has frozen into a single basin or glass form is incorrect.
The entropy reduction for $\Sigma$\ only happens if a microstate measurement
is performed to identify the particular component the $\Sigma$-glass is in.
This discussion also shows that calorimetric measurements explore different
glass components associated with the subsystems, so that they reveal a
non-zero residual entropy.

\subsection{Spontaneous Symmetry Breaking versus Glass Confinement}

It is relevant at this point to contrast the above component confinement in
glasses with the idea of confinement that occurs in spontaneous symmetry
breaking such as a ferromagnet. In the latter case, there exists a symmetry
breaking field, whose presence picks out one of the components, for example
$\lambda=\lambda_{0}$, for the system. This is equivalent to our notion of the
microstate measurement, except that in this case it is really a "component
measurement" resulting in picking the component $\lambda_{0}$: the application
of the symmetry breaking field forces the system to be in a particular
component $\lambda_{0}$. In this process, the entropy of the system will be
reduced by $S_{\mathcal{C}}$ (note that we have suppressed the time-dependence
as one usually considers equilibrium situations in symmetry-breaking) due to
the probability collapse discussed above on p. \pageref{marker_Prob_Collapse},
and all thermodynamic averages are determined by the particular component
$\lambda_{0}$. It usually happens that the value of $S_{\mathcal{C}}$ in
spontaneous symmetry breaking is not an extensive quantity, so the effects of
the confinement on the entropy become irrelevant for a macroscopic system.
There is no entropy reduction per unit volume in the process of confinement.
However, the effects of the confinement on some thermodynamic quantities, such
as the order parameter associated with the symmetry breaking, become very important.

For glasses, the residual entropy $S_{\mathcal{C}}(t)$ is found to be an
extensive quantity. Therefore, it is very important to know if there could be
some entropy reduction due to confinement in glasses. To this date, no one has
identified any physical symmetry breaking field analog for glasses. Thus,
there does not seem to be a way to prepare a glass in a particular component.
When a glass is prepared, we have no way of knowing which component it is
trapped in. Hence, one must consider all the components in obtaining any
thermodynamic average, such as the entropy as we have done above. In
particular, there is no entropy reduction per unit volume in a glass transition.

We have discussed the unsuitability of time-average (instead of the ensemble
average) elsewhere \cite{Gujrati-Symmetry}, and we refer the reader to this
for more details. We will only make the following brief comment here. It
happens that at low temperatures the approach to equilibrium takes more time
than feasible due to experimental constrained. Most measurements last a short
period of time. The temporal average over an extended time period has nothing
to do with information obtained in measurements that may take a fraction of a
second or so. Unless the system is already in equilibrium at the time of the
measurement, different measurements carried out on the system at different
instances will give different results; the system has a memory effect, when
the system is not in equilibrium. Thus, temporal average is not desirable for glasses.

In contrast, the ensemble average provides an instantaneous average and thus
bypasses the above objection of the finite measurement time.

\subsection{Role of Irreversibility on the Residual
Entropy\label{Marker_Irreversibility_Residual_Entropy}}

There is another way to understand the inequalities in Eq.
(\ref{Entropy_HeatCapacity}). In a vitrification experiment from a state A at
temperature $T_{0}$ in the supercooled liquid state which is still higher than
the glass transition temperature to the state A$_{0}$ at absolute zero, we
have along the path A$\rightarrow$A$_{0}$%
\begin{equation}
S(0)=S(T_{0})+%
%TCIMACRO{\tint \limits_{\text{A}}^{\text{A}_{0}}}%
%BeginExpansion
{\textstyle\int\limits_{\text{A}}^{\text{A}_{0}}}
%EndExpansion
d_{\text{e}}S+%
%TCIMACRO{\tint \limits_{\text{A}}^{\text{A}_{0}}}%
%BeginExpansion
{\textstyle\int\limits_{\text{A}}^{\text{A}_{0}}}
%EndExpansion
d_{\text{i}}S, \label{General_Entropy_Calculation}%
\end{equation}
where we have set $\Delta S_{\text{C}}=0$ as there is no latent heat in the
vitrification process, and where $dS=d_{\text{e}}S+d_{\text{i}}S$
\cite{Donder,deGroot,Prigogine,Guj-NE-I,Guj-NE-II}, with $d_{\text{i}}S\geq0$
representing the irreversible entropy generation and $d_{\text{e}}%
S=C_{P}dT_{0}/T_{0}$; see Eq. (\ref{Heat_Entropy Relation}). Since the second
integral in the above equation is always \emph{non-negative}, we obtain%
\begin{equation}
S(0)\geq S_{\text{expt}}(0)\equiv S(T_{0})+%
%TCIMACRO{\tint \limits_{T_{0}}^{0}}%
%BeginExpansion
{\textstyle\int\limits_{T_{0}}^{0}}
%EndExpansion
C_{P}dT_{0}/T_{0}, \label{Residual_Entropy_determination}%
\end{equation}
in accordance with Eq. (\ref{Entropy_HeatCapacity}), as expected. The forward
inequality is due to the irreversible entropy generation. Thus, the entropy at
absolute zero must be larger than or equal to the right hand side
$S_{\text{expt}}(0)$, which can be determined by performing a cooling
experiment. We take $T_{0}$ to be the melting temperature $T_{\text{M}},$ and
uniquely determine the entropy of the supercooled liquid at $T_{0\text{M}}$ by
adding the entropy of melting to the crystal entropy $S_{\text{CR}%
}(T_{0\text{M}})$ at $T_{0\text{M}}$. The latter is obtained in a unique
manner by integration along a reversible path from $T_{0}=0$ to $T_{0}%
=T_{0\text{M}}$:
\[
S_{\text{CR}}(T_{\text{M}})=S_{\text{CR}}(0)+%
%TCIMACRO{\tint \limits_{0}^{T_{\text{M}}}}%
%BeginExpansion
{\textstyle\int\limits_{0}^{T_{\text{M}}}}
%EndExpansion
C_{P\text{,CR}}dT_{0}/T_{0},
\]
here, $S_{\text{CR}}(0)$ is the entropy of the crystal at absolute zero, which
is traditionally taken to be zero, and $C_{P\text{,CR}}(T_{0})$ is the heat
capacity of the crystal. This then uniquely determines the entropy of the
liquid to be used in the right hand side in
Eq.(\ref{Residual_Entropy_determination}). If experiments show that
$S_{\text{expt}}(0)$ \emph{greater} than zero (or greater than $S_{\text{CR}%
}(0)$ if the latter is not taken to be zero in accordance with Nernst
postulate), then the entropy $S(0)$ at absolute zero on the left side itself
must be greater than zero (or greater than $S_{\text{CR}}(0)$). We will assume
that $S_{\text{CR}}(0)=0$. Thus, the experimental determination of the right
hand side of Eq. (\ref{Residual_Entropy_determination}) is sufficient to
unequivocally determine whether $S(0)\geq S_{\text{expt}}(0)$. The inequality
in Eq. (\ref{Residual_Entropy_determination}) takes into account any
irreversibility during vitrification. As is well known, $S_{\text{expt}}(0)$
is found to be non-negative, as discussed in Sect.
\ref{Marker_Irreversibility}. Thus, there cannot be any doubt that
\begin{equation}
S_{\text{R}}\equiv S(0)\geq S_{\text{expt}}(0); \label{Residual_Entropy_Bound}%
\end{equation}
experiments invariably give a non-zero value of $S_{\text{expt}}(0)$, which
then proves immediately that in thses cases, the residual entropy cannot vanish.

\section{Non-equilibrium Entropy and Gibbs Free energy
\label{Marker_Stability}}

\subsection{Consequences of the Second Law}

The important conclusions from the previous discussion can now be summarized
as follows. The conclusions are derived under the assumption that the entropy
$S_{0}$ of the isolated system is a continuous function of $E_{0},V_{0}%
,N_{0},\xi_{0}(t)$ and $t$.

\begin{enumerate}
\item[(A)] Medium under internal equilibrium

\begin{enumerate}
\item[1.] The Gibbs free energy $G$ of the system is continuous function of
its arguments and decreases continuously during relaxation toward its
equilibrium value.

\item[2.] The entropy $S$ of the system is continuous function of
$E(t),V(t),\xi(t)$ and $t$, which follows from Theorem
\ref{Theorem_Continuous_System_Entropy}.
\end{enumerate}

\item[(B)] Medium and system under internal equilibrium

\begin{enumerate}
\item[1.] The entropy $S$ of the system is continuous function of $E(t),V(t),$
and $\xi(t)$. This is a weaker form of the continuity condition, and follows
from the continuity of the Gibbs free energy in Theorem
\ref{Theorem_Gibbs_free_energy_continuity}. To see that we write%
\begin{equation}
S=(H-G)/T_{0}, \label{Continuous_Entropy_System}%
\end{equation}
and use the experimentally observed continuity of $H$ during the glass transition.

\item[2.] The entropy continuously decreases during relaxation toward its
equilibrium value according to Eq. (\ref{Entropy_variation_system}).
\end{enumerate}
\end{enumerate}

It follows from either (A2) or (B1) that the entropy cannot undergo any
discontinuity at a glass transition. Since the continuity of $G$ follows from
the second law, any discontinuous entropy reduction will result in the
violation of the second law, a point already made by Goldstein
\cite{Goldstein}.

It follows from (B2) that the behavior of entropy during relaxation only
supports Eq. (\ref{S_CV_Variation}) and not Eq. (\ref{S_ELV_Variation}). In
other words, only the conventional view of the entropy during component
confinement can be substantiated by the second law. However, this does not
prove that the Gibbs free energy $G_{\text{GL,UV}}(T_{0})$ is incorrect as its
variation in time conforms to the expected variation in Eq.
(\ref{Gibbs_Free_Energy_Variation}). The Gibbs free energy $G_{\text{GL,CV}%
}(T_{0})$\ also conforms to the expected variation in Eq.
(\ref{Gibbs_Free_Energy_Variation}). Does it mean that $G_{\text{GL,UV}}%
(T_{0})$ is just as acceptable as $G_{\text{GL,CV}}(T_{0})$ \ from the point
of view of the second law and non-equilibrium thermodynamics, even if
$S_{\text{GL,UV}}(T_{0})$ is not? Or is it possible that both Gibbs free
energies are unacceptable? Then what should be the correct form of the Gibbs
free energy for a glass? Is it possible that $G_{\text{GL,UV}}(T_{0})$ is
acceptable, even though $S_{\text{GL,UV}}(T_{0})$ is found to violate the
second law? Will this lead to some thermodynamic inconsistent in some way? Is
it possible that only $G_{\text{GL,CV}}(T_{0})$ is acceptable? We turn to this
issue now.

\subsection{Non-equilibrium Entropy evaluation}

The Gibbs free energy $G_{\text{GL,CV}}(T_{0})$ is shown by the short dashed
blue curve which gradually connects without any inflection point to
$G_{\text{SCL}}(T_{0})$, while $G_{\text{GL,UV}}(T_{0})$ is shown by the solid
red curve and by its discontinuous or rapid fall to $G_{\text{SCL}}(T_{0})$ in
Fig. \ref{Fig_Gibbs_Free_Energy}; the latter possesses an inflection point A
in the dashed red portion. We rule out the discontinuous jump in
$G_{\text{GL,UV}}(T_{0})$ as physically unacceptable because of (A1). Thus, we
only consider $G_{\text{GL,UV}}(T_{0})$ with a continuous fall to
$G_{\text{SCL}}(T_{0}).$

All known analyses of experimental data have been carried out under the
assumption of mechanical and thermal equilibrium. Therefore, we will also make
this common assumption in this section, according to which
\begin{equation}
A_{\text{S}}=A_{\text{V}}=0. \label{Standard_Assumption}%
\end{equation}
Accordingly, we will assume that the instantaneous temperature and pressure
$T(t),P(t)$ of the system are no different from the constant temperature and
pressure $T_{0},P_{0}$ of the medium. As there is no longer any need to
distinguish the two, we will use the common notation $T,P$ for both, with the
implicit assumption that they have no longer any $t$-dependence.

In our discussion below, we do not specifically endorse the statistical
formulation of the entropy in Eq. (\ref{Entropies}). We simply use the
thermodynamic notion of the entropy. The resulting differential of the Gibbs
free energy is now given by \cite{Gutzow-Book,Nemilov-Book}%
\begin{equation}
dG=-SdT+VdP-A_{\xi}d\xi. \label{Diff_GibbsFE}%
\end{equation}
The irreversible entropy generation is captured by the presence of $\xi$ in
the above equation. This form of $dG$ is exactly what is postulated in the
traditional non-equilibrium thermodynamics \cite{Donder,deGroot,Prigogine},
where the existence of non-equilibrium entropy is postulated. How it is
actually calculated is irrelevant for the general argument below. This point
should not be forgotten in the following discussion as the general conclusions
should be confirmed by any formulation of the entropy.

The entropy is given by
\begin{equation}
S\equiv-(\partial G/\partial T)_{P,\xi}, \label{S_G_Derivative}%
\end{equation}
which generalizes the conventional definition of the entropy for an
equilibrium system to a non-equilibrium system. The above relation shows that
we should be able to extract the entropy $S$ from the knowledge of the Gibbs
free energy $G$. Now, it is easy to show \cite{Nemilov-Book} that the heat
capacity is given by%
\begin{equation}
C_{P,\xi}(T)=T\left(  \partial S/\partial T\right)  _{P,\xi}=-T(\partial
^{2}G/\partial T^{2})_{P,\xi}\geq0; \label{Stability_Cond_Modified}%
\end{equation}
this is in accordance with Eq. (\ref{Heat_Capacity_Stability_1}). We can now
integrate Eq. (\ref{Stability_Cond_Modified}) to obtain the entropy in terms
of the heat capacity $C_{P,\xi}:$
\begin{equation}
S(T,\xi)=S(0,\xi)+%
%TCIMACRO{\tint \limits_{0}^{T}}%
%BeginExpansion
{\textstyle\int\limits_{0}^{T}}
%EndExpansion
\frac{C_{P,\xi}(T)dT}{T}+\Delta S_{\text{C}},
\label{Entropy_HeatCapacity_Identity}%
\end{equation}
where $S(0,\xi)$ is the residual entropy at absolute zero, and where
\[
dQ_{P,\xi}\equiv C_{P,\xi}(T)dT
\]
represents the amount of heat at constant $P,\xi$. The Eq.
(\ref{Entropy_HeatCapacity_Identity}) should be compared with Eq.
(\ref{Entropy_HeatCapacity0}), which is usually criticized as unreliable due
to the glass transition. In contrast, Eq. (\ref{Entropy_HeatCapacity_Identity}%
) is an identity in our non-equilibrium thermodynamics \cite{Guj-NE-I} under
the assumption in Eq. (\ref{Standard_Assumption}). Any irreversibility in the
system is now captured by the time dependence of $\xi$. However, the actual
form of the variation is not relevant for our discussion in this work. The
integration in this identity must be carried out along \emph{constant} $\xi$.
This heat capacity cannot vanish at any positive temperature. It can vanish at
absolute zero without creating any conceptual problem. We have discussed these
issues elsewhere \cite{Gujrati-Nernst,Gujrati-Fluctuations}. Any state with a
negative heat capacity will be identified here as \emph{unphysical}.

\subsection{Consequences of $\xi$ for Glasses}

Let us try to understand the consequence of the additional variable $\xi$. In
the presence of $\xi$, all the the curves including the broken red curve in
Figs. \ref{Fig_EntropyLoss} and \ref{Fig_Gibbs_Free_Energy}\ turn into
surfaces defined over the $T-\xi$ plane. In particular, the inflection point A
that is present in the dashed piece in this figure turns into a \emph{line
}A($\xi$)\emph{\ of inflection points}. Thus, if we consider a slice of these
surfaces at a constant $\xi=\xi_{\text{C}}$, then this slice will give three
curves similar to those shown in Figs. \ref{Fig_EntropyLoss} and
\ref{Fig_Gibbs_Free_Energy}, with $S_{\text{GL,UV}}(T_{0})$ and
$G_{\text{GL,UV}}(T_{0})$\ still containing inflection points. In particular,
we will find a region curving down at higher temperatures and a region curving
up at lower temperatures in Fig. \ref{Fig_EntropyLoss}, just as we see in the
figure for $S_{\text{GL,UV}}(T_{0})$. Nevertheless, the form of the entropy
$S_{\text{GL,CV}}(T_{0})$ or $S_{\text{GL,UV}}(T_{0})$\ for fixed $\xi
=\xi_{\text{C}}$ is \emph{monotonic} as shown in Fig. \ref{Fig_EntropyLoss},
so that
\begin{equation}
\left(  \frac{\partial S}{\partial T}\right)  _{P,\xi}=\frac{C_{P,\xi}(T)}%
{T}>0 \label{S_slope}%
\end{equation}
for both of them. Since we are allowed to take any value $\xi_{\text{C}}$ in
the above discussion, there is no harm in showing the fixed value as a general
$\xi$ in the above equation. The above inequality corresponds to $C_{P,\xi
}>0,$ in accordance with Eq. (\ref{Stability_Cond_Modified}). Again, we show
$\xi$ and not $\xi_{\text{C}}$\ for the reasons alluded to above. In an
isobaric experiment at fixed $P$, the curves shown in Fig.
\ref{Fig_EntropyLoss} need not correspond to a fixed $\xi$ in the glass
transition region. In general, we expect $\xi$ to vary with the temperature,
pressure and the history of cooling. But the point to stress is that the
presence of the inflection point A in Fig. \ref{Fig_EntropyLoss} has no effect
on the sign of the slope $\left(  \partial S/\partial T\right)  _{P,\xi}$. It
remains non-negative for both entropies $S_{\text{GL,CV}}(T_{0})$ or
$S_{\text{GL,UV}}(T_{0})$.

\subsection{Forms of Gibbs Free Energies\label{Marker_Inconsistency}}

The enthalpy $H_{\text{GL}}(T,\xi)$ is the same in both views%
\[
H_{\text{GL,CV}}(T,\xi)\equiv H_{\text{GL,UV}}(T,\xi)=H_{\text{GL}}(T,\xi),
\]
because of the continuity of the enthalpy during the glass transition. The
Gibbs free energy
\[
G_{\text{GL}}(T,\xi)\equiv H_{\text{GL}}(T,\xi)-TS_{\text{GL}}(T,\xi)
\]
thus, has the same value $H_{\text{GL}}(0,\xi)$ in CV and UV at absolute zero:%
\[
G_{\text{GL}}(0,\xi)=G_{\text{GL,CV}}(0,\xi)\equiv G_{\text{GL,UV}}%
(0,\xi)\equiv H_{\text{GL}}(0,\xi),
\]
regardless of the value $S_{\text{GL,CV}}(0,\xi)\ $and $S_{\text{GL,UV}}%
(0,\xi)$ take at absolute zero (we obviously do not consider the unphysical
case of an infinitely large $S(T,\xi)$ for a finite but macroscopically large
system), depending on whether we follow CV or UV. This is the same as Eq.
(\ref{Common_Gibbs_Free_Energy_at_0}). The SCL Gibbs free energy
$G_{\text{SLC}}(T)$ above $T_{\text{g}}\equiv T_{0\text{g}}$ is the same in
both views. If it is possible for the "Supercooled Liquid" curve to be
extrapolated to absolute zero, as shown by the solid blue curve in Fig.
\ref{Fig_Gibbs_Free_Energy}, then $G_{\text{SCL}}(T)$ would have the value
$G_{\text{SCL}}(0)=H_{\text{SCL}}(0)\leq H_{\text{GL}}(0,\xi)$ there. Observe
that%
\[
G_{\text{SLC}}(T)<G_{\text{GL,CV}}(T,\xi)<G_{\text{GL,UV}}(T,\xi)
\]
for $T<T_{\text{g}}$. The value $G_{\text{SCL}}(0)$ should be strictly lower
than $G_{\text{GL}}(0,\xi),$ as $G_{\text{GL,CV}}(T,\xi)$ or $G_{\text{GL,UV}%
}(T,\xi)$ is expected to reach $G_{\text{SCL}}(T)$ from above during
relaxation. The curve $G_{\text{SLC}}(T)$ presumably approaches absolute zero
with a slope equal to zero $(S_{\text{SCL}}(0)=0)$. The entire supercooled
liquid curve $G_{\text{SLC}}(T)$ is concave, as shown because it represents a
stable state.

\subsection{Constant internal order parameter}

Let us first consider the additional variable $\xi$ to be constant for the
sake of simplicity of the discussion. Therefore, we will suppress $\xi$ in
this section. At absolute zero, the Gibbs free energy in either view takes the
same value $G_{\text{GL}}(0)$, as$~$noted above; see the solid red curve and
the dashed blue curve in Fig. \ref{Fig_Gibbs_Free_Energy}. Thus, there are two
possible Gibbs free energy curves below the glass transition at $T_{\text{g}}%
$: $G_{\text{GL,UV}}(T)$ or $G_{\text{GL,CV}}(T)$. Since the entropy $S$ is
given by Eq. (\ref{S_G_Derivative}), it is evident that the magnitude of the
slope of $G_{\text{GL,UV}}(T)$ will be smaller $(\simeq0)$ than that of
$G_{\text{GL,CV}}(T)$ near absolute zero because of the entropy reduction, as
shown schematically. With no entropy reduction, $G_{\text{GL,CV}}(T)$ will
have a large, negative slope equal to $-S_{\text{R}}$ at absolute zero. Hence
it will drop faster than the upper red curve, so that $G_{\text{GL,CV}}(T)$
will be given by the short dashed blue curve in Fig.
\ref{Fig_Gibbs_Free_Energy}. This curve will eventually connect to the lower
solid blue curve $G_{\text{SLC}}(T)$ indicated by "Supercooled Liquid." The
combined curve (short dashed blue curve $G_{\text{GL,CV}}(T)$ below
$T_{\text{g}}$ and solid blue curve $G_{\text{SLC}}(T)$ above $T_{\text{g}}$)
remain \emph{concave} at all temperatures, as shown schematically in the
figure, as there is no inflection point. The Gibbs free energy
$G_{\text{GL,UV}}(T)$ must also be \emph{continuously} connected with
$G_{\text{SLC}}(T)$ along the long dashed red curve as shown in Fig.
\ref{Fig_Gibbs_Free_Energy}; now the connection give rise to an inflection
point A.

According to Eq. (\ref{Stability_Cond_Modified}), the Gibbs free energy must
be a concave function of the temperature so that the heat capacity remains
non-negative, as the solid red curve, the short dashed and solid blue curves
in Fig. \ref{Fig_Gibbs_Free_Energy} are. Remember that we are talking about
the total heat capacity of the system. It is clear that being concave,
$G_{\text{GL,CV}}(T)$\ gives a non-negative heat capacity. On the other hand,
this is not true of $G_{\text{GL,UV}}(T)$, which contains\ an inflection point
at A near $T_{\text{g}}$. It is clear that any attempt to \emph{smoothly}
connect the solid red piece $G_{\text{GL,UV}}(T)$ with $G_{\text{SLC}}(T)$
must result in an inflection, so that it cannot remain concave everywhere. It
must have a convex piece giving a negative heat capacity at higher
temperatures, as shown by the arrow. Let us emphasize that to connect the two
solid curves by a straight piece to avoid the convex piece is contrary to
physics. This will result in a discontinuity in the entropy \ and, hence, a
latent heat where the connection is made. This is contrary to all experimental
evidence. In addition, over this intermediate temperature range, the heat
capacity will turn out to be zero, which is also contrary to all experimental
evidence. Thus, the dashed piece must not be straight for a physically
realizable glass.

\subsection{Variable Internal order parameter}

Let us now understand the significance of Eq. (\ref{Stability_Cond_Modified})
when a variable internal order parameter $\xi$ is present. In this case, these
free energy curves will turn into surfaces defined over the $T_{0}-\xi$ plane.
In particular, the inflection point A in this figure turns into a \emph{line
of inflection point}, as is also the case with Fig. \ref{Fig_EntropyLoss}. Let
us consider a slice of these surfaces obtained by fixing $\xi=\xi_{0}$. As
this slice will also cut the line of inflection points, the slice will have
the middle piece in which the curvature will change sign on either side of
this inflection point. Thus, this slice will appear similar to the curves in
Fig. \ref{Fig_Gibbs_Free_Energy} with the middle dashed red curve containing
an inflection point at $T=T_{\text{A}}$. At temperatures lower than
$T_{\text{A}}$, the Gibbs free energy is concave so that Eq.
(\ref{Stability_Cond_Modified}) is satisfied. However, at temperatures higher
than $T_{\text{A}}$, the Gibbs free energy becomes convex so that Eq.
(\ref{Stability_Cond_Modified}) is not satisfied. This region then results in
a region of \emph{negative heat capacity}, and this makes the system
\emph{unphysical }over this region above $T_{0\text{A}}$. Indeed, all
experimental evidence collected so far for glasses have never ever exhibited a
negative heat capacity.

The presence of a point of inflexion at $T_{\text{A}}$, where the heat
capacity will vanish, raises another issue. It is obvious that $T_{\text{A}}$
is a positive temperature. This gives rise to another unphysical aspect of UV
in that the glass at this temperature can be brought in thermal equilibrium
with any medium at any temperature. No heat exchange is possible ($dQ_{P,\xi
}\equiv C_{P,\xi}dT=0$) as the system has vanishing heat capacity. This means
that at A where the heat capacity is zero, the temperature has no physical
significance for a glass; see the discussion above.

In addition to these problem, the negative heat capacity is in contradiction
with (\ref{S_slope}). The negative curvature of the Gibbs free energy
$G_{\text{GL,UV}}(T)$ implies that the entropy must be a decreasing function
of the temperature, in direct contradiction with the behavior of the entropy
$S_{\text{GL,UV}}(T)$ in Fig. \ref{Fig_EntropyLoss}. The entropy derived from
$G_{\text{GL,UV}}(T)$ is different from the entropy $S_{\text{GL,UV}}(T)$ in
Fig. \ref{Fig_EntropyLoss}. Thus, there is an \emph{internal inconsistency}, a
thermodynamic inconsistency, within the unconventional view. The lack of
thermodynamic consistency and the violation of Eq.
(\ref{Stability_Cond_Modified}) rules out the unconventional view as
physically relevant. Only the conventional view can be supported by the second law.

\section{Conclusions}

The formulation of the second law of thermodynamics in Eq. (\ref{Second_Law})
presupposes the existence and continuity of the entropy $S_{0}$ of an isolated
system under all possible conditions, and not only when the system is in
equilibrium. This point is, however, not always appreciated. A microstate is
identified by specifying some extensive mechanical quantities of the system
such as energy, volume, etc. that we have identified here as state variables
denoted collectively by $\mathbf{Z}(t)$. These variables, and therefore the
microstates and their probabilities $p_{i}(t)$ exists even if the fields like
the temperature, pressure etc. may not exist for the system. As the entropy is
a state function, the entropy $S_{0}$ as identified in Eq.
(\ref{Conventional_Entropy}) exists for an isolated body under all conditions,
and we have assumed it to be a continuous function of $\mathbf{Z}_{0}$ ant
$t$. This assumption must be made for Eq. (\ref{Second_Law}) to have any
content. We start from this and show that this results in the continuity of
the Gibbs free energy in Eq. (\ref{Free_Energies}) for an open system; see
Theorem \ref{Theorem_Gibbs_free_energy_continuity}. The Gibbs free energy
exists under all possible conditions of the system. In terms of the Gibbs free
energy, the second law for an open body is given by Eq.
(\ref{Gibbs_Free_Energy_Variation}).

If we now assume the open system to be in internal equilibrium, we can
identify its fields and affinities by Eq. (\ref{Field_Affinity}). For a body
under internal equilibrium, certain results have been obtained in Sect.
\ref{Marker_Useful_Results}, the most important being Eq.
(\ref{Relaxation_Facts}) and Eq. (\ref{Entropy_variation_system}) under
cooling when we have mechanical equilibrium ($P=P_{0}$). The last observation
is later used to rule out any theory of vitrification (via cooling) that
predicts an increase of entropy during relaxation. The statistical concept of
entropy is introduced in Sect. \ref{Marker_NonEq-S}, given by Gibbs
\cite{Gibbs} for an isolated system under all conditions. It follows form the
\textsc{Fundamental Axiom }in this section \cite{Gujrati-Symmetry}\textsc{.
}Under the assumption of quasi-independence, the entropy of an open system
follows from this statistical formulation and is given in Eq. (\ref{Entropies}).

The principles of non-equilibrium thermodynamics are enunciated in Sect.
\ref{Sect_Thermodynamic_Principles}, which follow from the \textsc{Fundamental
Axiom}. These principles reduce to accepted principles of equilibrium
thermodynamics. Using these principles, we explain how the entropy of a system
should remain continuous during component confinement at a glass transition.
We then prove the continuity of $S$ under two distinct situations; see (A2)
and (B1). This continuity of $S$ is one of the most important results of this
work and is derived under the most possible general condition; see Theorem
\ref{Theorem_Continuous_System_Entropy}. The statistical entropy merely
explains how this happens. The next most important result is captured by
Theorem \ref{Theorem_Entropy_Relaxation} according to which only Eq.
(\ref{S_CV_Variation}) can be justified by the second law; Eq
(\ref{S_ELV_Variation}) cannot be justified. The discussion also explains the
concept of the residual entropy. The residual entropy is estimated by the
experimentally obtained value $S_{\text{expt}}(0)$ of the entropy, see Eq.
(\ref{Residual_Entropy_determination}), by applying equilibrium thermodynamics
to the glasse, a standard practice in the field. The experimental estimate
$S_{\text{expt}}(0)$ is normally a non-negative value. We give a direct proof
that the residual entropy must be bounded from below by this experimental
value, see Eq. (\ref{Residual_Entropy_Bound}), thus making the residual
entropy at least as big as the experimental estimate. Any irreversibility will
only raise the value of the residual entropy even higher than the experimental
estimate. It is impossible for the residual entropy to be zero if the
experimenatl estimate $S_{\text{expt}}(0)$ is non-zero. This conclusion
justifies a non-zero $S_{\text{R}}$ at G in Fig. \ref{Fig_Gap_Model}.

Using general arguments of stability and well-established non-equilibrium
thermodynamics, we have also shown that $G_{\text{GL,UV}}(T)$ is internally
inconsistent in Sect. \ref{Marker_Inconsistency}. This conclusion is not based
on any particular formulation of the entropy. It only uses the thermodynamic
concept of entropy. Our conclusion then adds another argument in support of
the reality of the residual entropy. Any entropy reduction scenario during
component confinement violates the second law and results in an internal inconsistency.

We wish to thank Marty Goldstein for suggesting to consider the issue of the
residual entropy using a statistical mechanical approach and him and S.V.
Nemilov for useful correspondence.


\begin{thebibliography}{99}                                                                                               %


\bibitem {Guj-NE-I}P.D. Gujrati, Phys. Rev. E \textbf{81}, 051130 (2010); P.D.
Gujrati, arXiv:0910.0026.

\bibitem {Guj-NE-II}P.D. Gujrati, arXiv:1101.0438.

\bibitem {Gibbs}J.W. Gibbs, \textit{Elementary principles in statistical
mechanics}, Ox Bow Press,Woodbridge, Conn. (1981).

\bibitem {Landau}L.D.\ Landau, E.M. Lifshitz, \textit{Statistical Physics},
Vol. 1, Third Edition, Pergamon Press, Oxford (1986).

\bibitem {Huang}Huang, K., \textit{Statistical Mechanics}, Second Edition,
John Wiley and Sons, New York, USA (1987).

\bibitem {Rice}S.A. Rice and P. Gray, \textit{The Statistical Mechanics of
Simple Liquids}, Interscience Publishers, New York (1965).

\bibitem {Becker}Becker, R., \textit{Theory of Heat}, second edition revised
by G. Leibfried, Springer-Verlag, N.Y., USA (1967).

\bibitem {Gujrati-Symmetry}P.D. Gujrati, Symmetry \textbf{2}, 1201 (2010).

\bibitem {Donder}Th. de Donder and P. van Rysselberghe, \textit{Thermodynamic
Theory of Affinity}, Stanford University, Stanford (1936).

\bibitem {deGroot}S.R. de Groot and P. Mazur, \textit{Non-Equilibrium
Thermodynamics}\textbf{, }First Edition, Dover, New York (1984).

\bibitem {Prigogine}D. Kondepudi and I. Prigogine, \textit{Modern
Thermodynamics}, John Wiley and Sons, West Sussex (1998).

\bibitem {Palmer}R.G. Palmer, Philos. Mag. B \textbf{44}, 533 (1981); Adv.
Phys. \textbf{31}, 669 (1982).

\bibitem {Jackle}J. J\"{a}ckle, (a) Philos. Mag. B \textbf{44}, 533 (1981);
(b) Physica B \textbf{127}, 79 (1984).

\bibitem {Gujrati-book}P.D. Gujrati in \textit{Modeling and Simulation in
Polymers,} edited by P.D. Gujrati and A.I Leonov, Wiley-VCH, Weinheim (2010).

\bibitem {note0}See for example A.B. Bestul and S.S. Chang, J. Chem. Phys.
\textbf{43}, 4532 (1965).

\bibitem {Johari-New}G.P. Johari and J. Khouri, J. Chem. Phys. (to appear).

\bibitem {Jones}G.O. Jones, \textit{Glass}, Chapmann and Hall, Gateshead (1971).

\bibitem {GoldsteinSimha}\textit{The glass transition and the nature of the
glassy state}, edited by M. Goldstein and Robert Simha, N.Y. Acad. Sci., N.Y.(1976).

\bibitem {Nemilov-Book}S.V. Nemilov, Thermodynamic and Kinetic Aspects of the
Vitreous State, CRC Press, Boca Raton (1995).

\bibitem {Gutzow-Book}I. Gutzow and J.W.P. Schmelzer, \textit{The Vitreous
State: Thermodynamics, Structure, Rheology and Crystallization}, Springer,
Berlin (1995).

\bibitem {Gujrati-Nernst}P.D. Gujrati, Phys. Lett. A \textbf{151}, 375 (1990);
P.D. Gujrati, arXiv:cond-mat/0308439;

\bibitem {Gujrati-Fluctuations}P.D. Gujrati, Rec. Res. Devel. Chem. Physics,
\textbf{4}, 243 (2003).

\bibitem {GujratiResidualentropy}P.D. Gujrati, arXiv:0908.1075.

\bibitem {Gujrati-Defects}P.D. Gujrati, arXiv:0911.0649.

\bibitem {Pauling}L. Pauling and R.C. Tolman, J. Am. Chem. Soc. \textbf{47},
2148 (1925).

\bibitem {Tolman}R.C. Tolman, \textit{The Principles of Statistical
Mechanics},Oxford University, London (1959).

\bibitem {Chow}Y. Chow and F.Y. Wu, Phys. Rev. B \textbf{36}, 285 (1987); see
references in this work for other cases where the residual entropy is shown to
exist rigorously.

\bibitem {Giauque}W.E.F. Giauque and M. Ashley, Phys. Rev. \textbf{43}, 81 (1933).

\bibitem {Pauling-ice}L. Pauling, J. Am. Chem. Soc. \textbf{57}, 2680 (1935).

\bibitem {Nagle}J.F. Nagle, J. Math. Phys. \textbf{7}, 1484 (1966).

\bibitem {Isakov}S.V. Isakov, K.S. Raman, R. Moessner, and S.L. Sondhi, Phys.
Rev. B. \textbf{70}, 104418 (2004).

\bibitem {Berg}B.A. Berg, C. Muguruma, and Y. Okamoto, Phys. Rev. B.
\textbf{75}, 092202 (2007).

\bibitem {Speedy}R.K. Bowles and R.J. Speedy, Mole. Phys. \textbf{87}, 1349
(1996); ibid. \textbf{87}, 1671 (1996).

\bibitem {Hemmen}A.C.D. van Enter and J.L. van Hemmen, Phys. Rev. A
\textbf{29}, 355 (1984).

\bibitem {Thirumalai}D. Thirumalai, R.D. Mountain, and T.R. Kirkpatrick, Phys.
Rev. A \textbf{39}, 3563 (1989).

\bibitem {Kievelson}D. Kievelson, and H. Reiss, J. Phys. Chem. B \textbf{103},
8337 (1999).

\bibitem {Mauro-Gupta}J.C. Mauro, P.K. Gupta, and R.J. Loucks, J. Chem. Phys.
\textbf{126}, 184511 (2007).

\bibitem {Note1}See the recent issue of J. Non-Cryst. Solids, \textbf{355}
(2009) for various reports for and against this conjecture.

\bibitem {Gutzow}I. Gutzow and J.W.P. Schmelzer, J. Non-Cryst. Solids,
\textbf{355}, 581 (2009).

\bibitem {Nemilov}S.V. Nemilov, J. Non-Cryst. Solids, \textbf{355}, 607 (2009).

\bibitem {Goldstein}M. Goldstein, J. Chem. Phys. \textbf{128},154510 (2008).

\bibitem {Gujrati-Comments}P.D. Gujrati, arXiv:0909.0238;arXiv:0909.0734.

\bibitem {Johari}G.P. Johari, Thermochimica Acta \textbf{500}, 111 (2010).

\bibitem {Kozliak}E. Kozliak and F.L. Lambert, Entropy, \textbf{10}, 274 (2008).

\bibitem {Keenan}J.H. Keenan, \textit{Thermodynamics}, MIT Press, Cambridge (1941).

\bibitem {Langer}E. Bouchbinder and J.S. Langer, Phys. Rev. E \textbf{80},
031132 (2009).

\bibitem {Onsager}L. Onsager, Phys. Rev.\textbf{37}, 405 (1931); \textbf{38},
2265 (1931)).

\bibitem {Jaynes}E.T. Jaynes, \textit{Papers on Probability, Statistics and
Statistical Physics}, R.D. Resenkrantz, Ed., reidel Publishing, Dordrecht,
Hollad (1983).
\end{thebibliography}
\end{document}